\documentclass[11pt]{article}
\usepackage{epsfig,amsmath,amsfonts,amssymb,amstext,afterpage,psfrag}

\usepackage{color}
\usepackage[english]{babel} 
\usepackage[utf8]{inputenc}
\usepackage[left=01in,right=1in,top=1.0in,bottom=1.1in]{geometry}
\usepackage[labelfont=bf]{caption}
\usepackage[hyperfootnotes=false]{hyperref}
\usepackage{color,amsmath,amssymb,exscale,psfrag,epsfig}
\usepackage{amsmath,amssymb,bm,slashed,array,bbold}
\usepackage{cite,color,url}
\usepackage{pdfpages}
\usepackage{graphicx}
\usepackage{slashed}
\usepackage[T1]{fontenc}
\usepackage[dvipsnames]{xcolor}
\usepackage{accents}
\usepackage{adjustbox}
\usepackage{multirow}
\usepackage{bm}
\usepackage{booktabs}
\usepackage{placeins}

\newcommand{\ifb}{\text{ fb}^{-1}}

\allowdisplaybreaks

\title{Indirect bounds on axion-like particle interactions from the SMEFT}

\begin{document}
\begin{titlepage}

\begin{flushright}
\normalsize
MITP-23-037
\end{flushright}

\vspace{3mm}
\begin{center}
\huge\bf
A global analysis of axion-like particle interactions using SMEFT fits
\end{center}

\vspace{-1mm}
\begin{center}
Anke Biekötter$^a$, Javier Fuentes-Martín$^b$, Anne Mareike Galda$^a$ and Matthias Neubert$^{a,c}$\\
\vspace{0.7cm} 
{\sl ${}^a$PRISMA$^+$ Cluster of Excellence \& Mainz Institute for Theoretical Physics\\
Johannes Gutenberg University, 55099 Mainz, Germany\\[2mm]
${}^b$Departamento de Física Teórica y del Cosmos, Universidad de Granada,\\ Campus de Fuentenueva, E–18071 Granada, Spain}\\[2mm]
${}^c$Department of Physics, LEPP, Cornell University, Ithaca, NY 14853, U.S.A.\\[2mm]
\end{center}

\vspace{0.8cm}
\abstract{In the presence of an axion or axion-like particle (ALP) that couple to the Standard Model via dimension-five interactions, dimension-six SMEFT interactions are generated via renormalization-group evolution. As many of these SMEFT contributions are experimentally tightly constrained, this ``ALP\,--\,SMEFT interference'' can be used to derive indirect bounds on the ALP couplings to the Standard Model particles. We present a global analysis of the Wilson coefficients of the ALP effective Lagrangian based on Higgs, top, and low-energy data. The obtained bounds are model independent and are competitive or even stronger than direct bounds in the GeV to TeV ALP-mass range.}
\end{titlepage}
\tableofcontents

\section{Introduction}

The lack of new-particle discoveries at the Large Hadron Collider (LHC) other than the Higgs boson restricts the space of possibilities for physics beyond the Standard Model~(SM). To have evaded detection so far, new particles must either be too heavy to be produced in current high-energy experiments, or interact very weakly with the SM particles. Pseudo Nambu–Goldstone bosons that appear from the spontaneous breaking of a global symmetry, commonly referred to as axions and axion-like particles (ALPs), are among the best motivated light particles beyond the SM. They could play a crucial role in the solution to the strong CP~problem~\cite{Peccei:1977hh,Weinberg:1977ma,Wilczek:1977pj} and are potential dark matter candidates (see~\cite{Chadha-Day:2021szb} for a review). While most explicit models addressing the strong CP~problem predict a strict relation between the axion mass and its decay constant, it is also possible to obtain solutions to the strong CP problem with heavier axions~\cite{Holdom:1982ex,Flynn:1987rs,Rubakov:1997vp,Berezhiani:2000gh,Hook:2014cda,Fukuda:2015ana,Gherghetta:2016fhp,Dimopoulos:2016lvn,Agrawal:2017ksf,Gaillard:2018xgk,Fuentes-Martin:2019bue,Gherghetta:2020keg,Kivel:2022emq}. Furthermore, composite-Higgs models~\cite{Gripaios:2009pe,Ferretti:2013kya} and even supersymmetric extensions of the SM~\cite{Nelson:1993nf,Bagger:1994hh,Bellazzini:2017neg} naturally give rise to ALPs. These models provide a good motivation to search for light pseudoscalar particles in a parameter space that extends beyond the traditional axion searches. Indeed, one of the main interests in searching for ALPs is that they can be a forerunner of a high-scale new physics sector, which otherwise could be hard to access experimentally.

Direct ALP searches span from cosmological~\cite{Cadamuro:2011fd, Millea:2015qra} and astrophysical~\cite{Payez:2014xsa,Jaeckel:2017tud} observations to collider~\cite{Mimasu:2014nea,Jaeckel:2015jla,Knapen:2016moh, Brivio:2017ije, Bauer:2017ris,Bauer:2018uxu } and flavor~\cite{Batell:2009jf,Freytsis:2009ct,Dolan:2014ska,MartinCamalich:2020dfe,Bauer:2019gfk, Bauer:2021mvw} experiments. These observables impose powerful constraints on the ALP parameter space, although often under very specific assumptions on its interactions. For example, direct searches for ALPs at particle colliders must make an assumption about the process in which the ALP is produced (e.g.\ in the decay of a Higgs boson, and $Z$ boson, or in weak decays of kaons or $B$ mesons), about its lifetime, and the way in which it decays, which may involve a decay outside of the detector. In the derivation of astrophysical bounds on, e.g., the ALP--photon coupling from supernovae or ALPs produced in the sun, it makes a crucial difference whether or not a vanishing ALP--electron coupling is assumed. In the vast majority of the existing analyses, it was assumed that only a single ALP couplings is non-zero at the scale of Peccei--Quinn symmetry breaking -- an assumption which is not valid in even the classic KSVZ \cite{Kim:1979if,Shifman:1979if} and DFSZ \cite{Dine:1981rt,Zhitnitsky:1980tq} models for the QCD axion.

As recently shown in~\cite{Galda:2021hbr}, the presence of ALP couplings to SM particles yields a non-trivial renormalization-group (RG) flow into the Wilson coefficients of the dimension-six Standard Model Effective Field Theory (SMEFT) operators. This opens up the interesting possibility of indirectly testing ALP interactions by utilizing existing SMEFT studies. In this paper, we exploit the ALP\,--\,SMEFT interference to deduce constraints from a global fit including low-energy, Higgs and top data, as implemented in current global SMEFT analyses~\cite{Falkowski:2017pss,Biekotter:2018ohn,daSilvaAlmeida:2018iqo,Brown:2018gzb,Hartland:2019bjb,Brivio:2019ius,Kraml:2019sis,vanBeek:2019evb,DeBlas:2019ehy,Dawson:2020oco,Almeida:2021asy,Ellis:2020unq,Anisha:2021hgc,Ethier:2021bye,Iranipour:2022iak,Bruggisser:2022rhb,Breso-Pla:2023tnz,Kassabov:2023hbm,Grunwald:2023nli},
thus providing a complementary use of these observables to constrain light new physics. As we show, our analysis probes previously unexplored parts of the ALP parameter space, especially for ALP masses above 10\,GeV. Furthermore, one of the significant advantages of our indirect approach is that the derived bounds are mostly independent of the ALP mass and require no assumptions on other ALP couplings or the ALP lifetime. This feature overcomes one of the main limitations of direct ALP searches, which often produce highly model-dependent constraints. Therefore, the indirect ALP bounds presented in this paper offer an important complementary approach with respect to direct constraints even when the latter appear to be stronger under certain conditions. 

This paper is organized as follows: Section~\ref{sec:ALPCouplings} introduces the ALP\,--\,SMEFT Lagrangian and discusses its connection to the SMEFT. In Section~\ref{sec:global_analysis}, we describe the global analysis setup, present the fit results and compare them to bounds from direct searches. Section~\ref{sec:ALP_UV} is dedicated to a reinterpretation of our global analysis in terms of concrete ultraviolet (UV) completions. We conclude in Section~\ref{sec:conclusion}. 

\section{ALP couplings to the SM}
\label{sec:ALPCouplings}

We consider an extension of the SM that includes a pseudoscalar, gauge-singlet ALP as an additional light state.  Its couplings to SM fields are, at the classical level, protected by an approximate shift symmetry $a\to a+c$, broken softly by the ALP mass term $m_a$. We relegate the discussion of possible UV completions of this model to Section~\ref{sec:ALP_UV}.

\subsection{The ALP Lagrangian}

It has been pointed out in \cite{Galda:2021hbr} that a consistent effective field theory (EFT) for an extension of the SM featuring an ALP must necessarily include higher-dimensional operators built out of SM fields only.
Above the electroweak scale, the most general effective Lagrangian describing the interactions of an ALP with SM particles thus reads
\begin{equation}
\label{eq:ALP_lag}
   \mathcal{L} = \frac{1}{2}\,(\partial_\mu a ) (\partial^\mu a ) - \frac{m_a^2}{2}\,a^2 + \mathcal{L}_\mathrm{SM+ALP} + \mathcal{L}_\mathrm{SMEFT} \, ,
\end{equation}
with $\mathcal{L}_\mathrm{SMEFT}$ denoting the SMEFT Lagrangian. The SMEFT Lagrangian up to dimension-six order can be found in~\cite{Grzadkowski:2010es}. In what follows, we adopt the same conventions as in this reference, except for the labels for the Higgs field and its (tachyonic) mass, for which we use $H$ and $m_H$, respectively. Note, in particular, that we work with dimensionless Wilson coefficients and factor out inverse powers of the new-physics scale $\Lambda$ when writing down higher-dimensional operators in the effective Lagrangian. The Lagrangian $\mathcal{L}_\mathrm{SM+ALP}$ describes the ALP interactions with SM particles. Apart from the soft symmetry breaking by the ALP mass $m_a,$ we consider only classically shift-invariant interactions with the SM fields. Specifically, the ALP Lagrangian up to dimension-six operators is then given by\footnote{We do not include redundant operators such as $\partial^\mu a\,(H^\dagger i\overleftrightarrow
 D_\mu H)$, which can be rewritten in terms of the ones here by means of field redefinitions~\cite{Bauer:2020jbp}.}
\begin{align}
\label{eq:lag1}
   \mathcal{L}_\mathrm{SM+ALP}^{D\le 6}=&  \,c_{GG}\,\frac{a}{f}\,\frac{\alpha_s}{4\pi}\,G_{\mu\nu}^A\,\tilde G^{\mu\nu\,A} + c_{WW}\,\frac{\alpha_L}{4\pi}\,\frac{a}{f}\,W_{\mu\nu}^I\,\tilde W^{\mu\nu\,I} + c_{BB}\,\frac{\alpha_Y}{4\pi}\,\frac{a}{f}\,B_{\mu\nu}\,\tilde B^{\mu\nu} \nonumber\\
   & + \frac{\partial^\mu a}{f}\,\sum_F \bar F\, \bm{c}_F \,\gamma_\mu\, F + \frac{c_{HH}}{f^2}\,(\partial^\mu a)(\partial_\mu a)\,H^\dagger H\,,
\end{align}
where $\bm{c}_F$ ($F=q,u,d,l,e$) are $3\times3$ hermitian matrices and $c_{ii}$ ($i=G,W,B,H$) are real couplings. The real parameter $f$ is referred to as the ALP decay constant, which is related to the relevant new-physics scale by 
\begin{align}
   \Lambda = 4\pi f \,.
\end{align}
The Lagrangian above can be cast into an alternative but equivalent form, in which the ALP couplings to fermions are of non-derivative type. This is achieved by means of the field redefinitions $F \to F + i\,\frac{a}{f}\,\bm{c}_F F$ in \eqref{eq:ALP_lag}, which yields
\begin{align}\label{eq:lag2}
\mathcal{L}_\mathrm{SM+ALP}^{D\le 6}&\to C_{GG}\,\frac{a}{f}\,G_{\mu\nu}^A\,\tilde G^{\mu\nu\,A} + C_{WW}\,\frac{a}{f}\,W_{\mu\nu}^I\,\tilde W^{\mu\nu\,I} + C_{BB}\,\frac{a}{f}\,B_{\mu\nu}\,\tilde B^{\mu\nu} \nonumber\\
&\quad - \frac{a}{f} \left( \bar{q} \, \tilde{H} \,\tilde{\bm{Y}}_u \, u_R + \bar{q} \, H \,\tilde{\bm{Y}}_d \, d_R + \bar{l} \, H \,\tilde{\bm{Y}}_e \, e_R + \mathrm{h.c.}\right) \nonumber\\
&\quad + \frac{1}{2}\,\frac{a^2}{f^2}\,\left(\bar q\,\tilde{H}\,\bm{Y}_u^\prime \,u_R +\bar q\,H\,\bm{Y}_d^\prime \,d_R +\bar l\, H\, \bm{Y}_e^\prime \,e_R + \mathrm{h.c.}\right)+ \frac{c_{HH}}{f^2}\,(\partial^\mu a)(\partial_\mu a)\,H^\dagger H\,.
\end{align}
Note that due to the axial anomaly, additional contributions enter the ALP couplings to gauge bosons, such that $c_{VV}$ and $C_{VV}$ are related by \cite{Bauer:2020jbp}
\begin{align}
\label{eq:related_coefs_bosons}
    C_{GG} &= \frac{\alpha_s}{4\pi}\,\left[ c_{GG} + \frac{1}{2}\,\text{Tr}(\bm{c}_d  +\bm{c}_u - 2\bm{c}_q) \right] , \notag\\
    C_{WW} &= \frac{\alpha_L}{4\pi}\,\left[ c_{WW} - \frac{1}{2}\,\text{Tr}(N_c\,\bm{c}_q  +\bm{c}_l) \right] ,\nonumber\\
    C_{BB} &= \frac{\alpha_Y}{4\pi}\,\left[ c_{BB} +\text{Tr}\Big(N_c \,(\mathcal{Y}_d^2\,\bm{c}_d + \mathcal{Y}_u^2\,\bm{c}_u - 2 \,\mathcal{Y}_q^2\,\bm{c}_q)  + \mathcal{Y}_e^2\,\bm{c}_e - 2\,\mathcal{Y}_l^2\,\bm{c}_l \Big) \right] ,
\end{align}
with $N_c = 3$ and the hypercharges $\mathcal{Y}_u = 2/3,\,\mathcal{Y}_d = -1/3,\, \mathcal{Y}_q = 1/6,\,\mathcal{Y}_e = -1,\,\mathcal{Y}_L = -1/2$.
The new ALP--fermion coupling matrices are related to the original Lagrangian couplings by
\begin{align}
\tilde{\bm{Y}}_{F_R} &= i\, (\bm{Y}_{F_R} \,\bm{c}_{F_R} - \bm{c}_{F_L}\, \bm{Y}_{F_R} )\,, \notag\\
\bm{Y}_{F_R}^\prime &= \bm{c}_{F_L}^2\, \bm{Y}_{F_R} - 2\, \bm{c}_{F_L}\,\bm{Y}_{F_R}\,\bm{c}_{F_R} + \bm{Y}_{F_R}\,\bm{c}_{F_R}^2\,,
\end{align} 
with $F_R=u,d,e$ and $F_L=q\,(l)$ for quark (lepton) couplings. Assuming a flavor-universal structure for the $\bm{c}_F$ couplings, i.e. $\bm{c}_F = c_F\, \mathbb{1}_3$, these interactions reduce to
\begin{align} \label{eq:ALP_Yukawa}
\tilde{\bm{Y}}_{F_R} = i \,\bm{Y}_{F_R} \, C_{F_R}\,, \qquad
\bm{Y}_{F_R}^\prime = \,\bm{Y}_{F_R} \, C_{F_R}^2\,, 
\end{align}
where $C_{u,d}\equiv c_{u,d}-c_q$ and $C_e\equiv c_e-c_l$. In the flavor-universal scenario, dimension-five ALP interactions with SM particles are thus fully described by six free parameters: $C_{GG}$, $C_{WW}$, $C_{BB}$, $C_u$, $C_d$ and $C_e$.\footnote{Given the hierarchical structure of the SM Yukawas, similar leading-order interactions would also be obtained for other flavor hypotheses in which third-family interactions are not (strongly) suppressed. The fermionic couplings $C_{u,d,e}$ correspond to the couplings $c_{ff}$ in~\cite{Galda:2021hbr} in flavor-universal scenarios, e.g.\ $C_u = c_{uu} =c_{cc} = c_{tt}$.}

\subsection{\texorpdfstring{ALP\,--\,SMEFT}{ALP-SMEFT} interference}

The dimension-five ALP couplings to SM particles generate divergent amplitudes that require dimension-six SMEFT counterterms. As a result, the RG equations for the SMEFT operators get modified by additional inhomogeneous source terms, such that the scale dependence in the presence of the ALP is given by 
\begin{equation}
    \frac{d}{d\ln\mu}\,C^\mathrm{SMEFT}_i - \gamma_{ji}^{\rm SMEFT}\,C^\mathrm{SMEFT}_j 
    = \gamma^{\rm SMEFT-ALP}_{i\alpha\beta}\,C^{\rm ALP}_\alpha\,[C^{\rm ALP}_\beta]^*\, .
\end{equation}
Here $\gamma^{\rm SMEFT}$ denotes the anomalous-dimension matrix of the dimension-six SMEFT operators calculated in~\cite{Elias-Miro:2013mua,Jenkins:2013zja,Jenkins:2013wua,Alonso:2013hga}, and the source terms $\gamma^{\rm SMEFT-ALP}$ have been calculated in~\cite{Galda:2021hbr}.\footnote{We follow the same conventions as in~\cite{Grzadkowski:2010es}, implying a factor of $1/2$ difference in the definition of $\lambda$ when compared to~\cite{Jenkins:2013zja,Jenkins:2013wua,Alonso:2013hga}, and a sign difference in the covariant derivatives relative to~\cite{Galda:2021hbr}.} The presence of ALP interactions with the SM particles thus generates a non-trivial RG flow into the SMEFT Wilson coefficients. The dimension-six inhomogeneous source terms are independent of the ALP mass. Thus, as long as the ALP mass lies below the scale of the observables used in the fit, we will obtain mass-independent indirect bounds on the ALP couplings. 

In addition to the contributions to the RG equations of the dimension-six SMEFT operators, the presence of the ALP also introduces modifications to the running of dimension-four couplings. Most of these modifications were discussed in~\cite{Galda:2021hbr}. However, additional contributions arise from the dimension-six operators included in~\eqref{eq:lag2}, which we present here for the first time. They are
\begin{align}
\label{eq:SM_ALP_running}
\frac{d\bm{Y}_{u,d,e}}{d\ln\mu} \supset -\frac{m_a^2}{2\Lambda^2}\, \bm{Y}_{u,d,e}^\prime\,, \qquad
\frac{dm_H^2}{d\ln\mu} \supset \frac{m_a^4}{\Lambda^2}\,c_{HH}\,,
\end{align}
where again $\Lambda=4\pi f$.

\subsection{Solving the \texorpdfstring{ALP\,--\,SMEFT}{ALP-SMEFT} RG equations}
\label{sec:RG_solution}

The RG equations for the ALP\,--\,SMEFT Lagrangian can generically be written as\footnote{The anomalous-dimension matrix $\gamma^{(5)}$ has been calculated in \cite{Chala:2020wvs,Bauer:2020jbp,Bonilla:2021ufe}. Numerical solutions to the corresponding RG equations can be obtained with the \texttt{ALPRunner} package~\cite{DasBakshi:2023lca}.} 
\begin{align}\label{eq:GenericRGEs}
\frac{d C^{(4)}_a(t)}{dt}&=\gamma_{ba}^{(4)}\big({\bm C^{(4)}},{\bm C^{(5)}},{\bm C^{(6)}}\big)\,C_b^{(4)}(t)\,,\notag\\
\frac{d C^{(5)}_\alpha(t)}{dt}&=\gamma_{\beta\alpha}^{(5)}\big({\bm C^{(4)}}\big)\,C_\beta^{(5)}(t)\,,\nonumber\\
\frac{d C^{(6)}_i(t)}{dt}&=\gamma_{ji}^{(6)}\big({\bm C^{(4)}}\big)\,C_j^{(6)}(t)+ \gamma^{(5,5)}_{i\alpha\beta}\big({\bm C^{(4)}}\big)\,C_\alpha^{(5)}(t)\,[C_\beta^{(5)}(t)]^*\,,
\end{align} 
where $t\equiv\ln\mu$, and we have collected the Wilson coefficients into vectors $\bm C^{(D)}$, with the superscript denoting the dimension of the associated operator, and the indices $a,b$ (for $D=4$), $\alpha,\beta$ (for $D=5$), and $i,j$ (for $D=6$) labeling the corresponding coefficients. $\gamma^{(D)}$ are the corresponding anomalous-dimension matrices, and $\gamma^{(5,5)}$ is the tensor accounting for the ALP source terms. The RG equations above form a system of coupled differential equations, for which an analytical solution is not known. One would thus be forced to solve this system numerically for a given set of initial conditions. It helps, however, to note that the RG equations above contain terms that are beyond dimension-six order in the EFT expansion. Indeed, using that ${\bm C^{(4)}}(t)={\bm C^{\rm SM}}(t)+\mathcal{O}(\Lambda^{-2})$, with ${\bm C^{\rm SM}}$ being the SM couplings, we can rewrite the RG equations for the Wilson coefficients of the higher-dimensional operators in the form
\begin{align}
\frac{d C^{(5)}_\alpha(t)}{dt}&=\gamma_{\beta\alpha}^{(5)}\big({\bm C^{\rm SM}}\big)\,C_\beta^{(5)}(t) \,,\nonumber\\
\frac{d C^{(6)}_i(t)}{dt}&=\gamma_{ji}^{(6)}\big({\bm C^{\rm SM}}\big)\,C_j^{(6)}(t)+ \gamma^{(5,5)}_{i\alpha\beta}\big({\bm C^{\rm SM}}\big)\,C_\alpha^{(5)}(t)\,[C_\beta^{(5)}(t)]^* \,,
\end{align}
in which all power-suppressed terms have been expanded out consistently. This system of equations admits an analytic solution for ${\bm C^{(5,6)}}(t)$ in terms of ${\bm C^{\rm SM}}(t)$.\footnote{Note that the running of $\bm C^{(4)}$ up to dimension-six order still needs to be determined numerically for each set of initial conditions. However, the system of equations to be solved is now considerably smaller.} Explicitly, we find
\begin{align}\label{eq:Evolution}
C^{(5)}_\alpha(t_f)&=U^{(5)}_{\alpha\beta}(t_f,t_0)\,C^{(5)}_\beta(t_0)\,, \notag\\
C^{(6)}_i(t_f)&=U^{(6)}_{ij}(t_f,t_0)\,C^{(6)}_j(t_0)+U^{(5,5)}_{i\alpha\beta}(t_f,t_0)\,C^{(5)}_\alpha(t_0)\,[C^{(5)}_\beta(t_0)]^* \,,
\end{align}
where $t_0$ and $t_f$ denote the logarithms of the initial and final energy scales, respectively, and the evolution tensors are defined as
\begin{align}\label{eq:U55solu}
U^{(D)}(t_f,t_0)&\equiv\mathcal{T} \exp\left[\int_{t_0}^{t_f}\,\big[\gamma^{(D)}\big({\bm C^{\rm SM}}(w)\big)\big]^T\,dw\right] ,\nonumber\\
U^{(5,5)}_{i\alpha\beta}(t_f,t_0)&\equiv U^{(6)}_{ij}(t_f,t_0)\int_{t_0}^{t_f} U^6_{jk}(t_0,w)\,\gamma^{(5,5)}_{k\rho\sigma}\big({\bm C^{\rm SM}}(w)\big)\,U^{(5)}_{\rho\alpha}(w,t_0)\,[U^{(5)}(w,t_0)_{\sigma\beta}]^*\,,
\end{align}
with $\mathcal{T}$ indicating that the exponential is $t$-ordered. Obtaining the evolution tensors is computationally expensive;  however, they do not depend on the initial conditions set on the higher-dimensional operators and only need to be determined once (for a given set of SM input parameters). Once the evolution tensors are known, the computation of the running of the Wilson coefficients for arbitrary initial conditions is very fast. 

The evolution matrix $U^{(6)}$ for the SMEFT has already been determined and is part of the computer tool \texttt{DsixTools} 2.0~\cite{Celis:2017hod,Fuentes-Martin:2020zaz}. We have used a customized version of \texttt{DsixTools}, in which we have incorporated the ALP contributions to the RG equations, to compute the evolution tensor $U^{(5,5)}$ in the flavor-universal scenario. We provide this tensor for the SMEFT Wilson coefficients relevant for our fits in a \texttt{Mathematica} notebook as ancillary material. A new version of \texttt{DsixTools} featuring generic ALP contributions will be presented in a forthcoming paper. 

\section{Global analysis of ALP couplings}
\label{sec:global_analysis}

In the following, we utilize the ALP\,--\,SMEFT interference to derive (almost) mass-independent bounds on the ALP couplings to SM particles. Contrary to the direct bounds derived elsewhere, the results we obtain are model-independent in the framework we consider, consisting of an ALP added to the SM without additional sources of new physics. One of the advantages of this approach is that it lets us efficiently reuse results from existing SMEFT analyses. 

\subsection{Experimental inputs and SMEFT predictions}
\label{sec:fit_input}

Limits on the SMEFT Wilson coefficients have been derived from a multitude of observables, including low-energy, Higgs and top data. For our global analysis, we utilize existing parametrizations of these observables in terms of the SMEFT Wilson coefficients and recast them in terms of ALP Wilson coefficients at the high-energy scale $\Lambda$. All of the SMEFT predictions used here are truncated at linear order in the dimension-six Wilson coefficients and employ $\{G_F , \, \alpha , \, M_Z \}$ as the electroweak input parameters. 

Predictions for the Higgs sector are taken from \cite{Anisha:2021hgc} and references therein, while SMEFT predictions for the top sector are taken from \texttt{fitmaker}~\cite{Ellis:2020unq} and references therein. The experimental observables used for Higgs and top data are summarized in Tables~\ref{tab:obset}-\ref{tab:obset_top2} in Appendix~\ref{app:exp_inputs}. The assumption of flavor universality in some SMEFT parametrizations of these observables, such as the ones from the experimental analysis in~\cite{ATLAS:2020naq}, causes certain complications, as ALP contributions are generally flavor-dependent. To overcome this issue, we will assume that the effects from operators involving quark couplings to gluons or the Higgs boson are dominated by those involving third-generation quarks, i.e.\
\begin{align}
    C_{uG} \to [C_{uG}]_{33} \,, \qquad
    C_{uH} \to [C_{uH}]_{33} \,, \qquad
    C_{dH} \to [C_{dH}]_{33} \,, 
\end{align}
where the notation $[C_{x}]_{ij}$ is used to denote the flavor indices $i$ and $j$ of the Wilson coefficient $C_x$.
The constraints on the remaining operators with flavor indices are typically experimentally dominated by first- and second-generation couplings. We therefore take a conservative approach and assume pure second-generation contributions for these operators in the considered Higgs and top observables, e.g.\ $C^{(3)}_{Hl}\to [C^{(3)}_{Hl}]_{22}$.

We perform a $\chi^2$ fit for the experimental data $\Vec{d}$ with the theory predictions $\Vec{p}(C_i)$ and covariance matrix $\bm{V}$, using the definition 
\begin{align}
    \chi^2(C_i) = \left[\Vec{d} - \Vec{p}(C_i)\right]^T \bm{V}^{-1} \left[\Vec{d} - \Vec{p}(C_i)\right] .
\end{align}
For low-energy observables including electroweak precision data, neutrino scattering, atomic parity violation and quark pair-production at LEP2, we directly use the $\chi^2$~function provided in~\cite{Efrati:2015eaa,Falkowski:2017pss}, which includes the full flavor structure of the SMEFT Wilson coefficients.\footnote{Notice that the definition of $C_{ll}$ and $C_{ee}$ in these references differs by a factor $1/2$ from the usual Warsaw basis definition. We have rescaled the corresponding Wilson coefficients to match the standard definitions employed for the Higgs and top sectors.} 

The SMEFT predictions are translated to ALP predictions using the solution to the RG equations in \eqref{eq:Evolution} for $C_i^{(6)}(t_0)=0$, i.e.\ neglecting possible matching contributions at the high scale. The value of the high-scale is set to $\mu_0=\Lambda=4\pi f$, and the low-energy scale is identified with the relevant experimental scale for each of the observables, $\mu_f=\mu_{\rm exp}$, with e.g.\ $\mu_\text{exp}=m_h$ for gluon-fusion Higgs production observables and $\mu_\text{exp}=m_h+2 m_t$ for $t\bar{t}h$~production. When considering ALP masses above the $Z$ mass, as we will do in Section~\ref{sec:comparison_with_direct_bounds}, we stop the ALP-induced running at $m_a$ and use the pure SMEFT running below this scale. ALP contributions to the RG evolution of dimension-four parameters such as the top-quark Yukawa coupling $y_t$ in \eqref{eq:SM_ALP_running} are found to be numerically irrelevant for the present analysis and have therefore been neglected. 

The total $\chi^2$ function is obtained by adding the individual contributions from low-energy observables, Higgs observables, and top data. The $\chi^2$ functions for all these data sets, both written in terms of SMEFT Wilson coefficients and ALP parameters, are provided in the ancillary material.

\subsection{Fit results}\label{sec:fit_results}

\begin{figure}
    \centering
    \includegraphics[width=.6\textwidth]{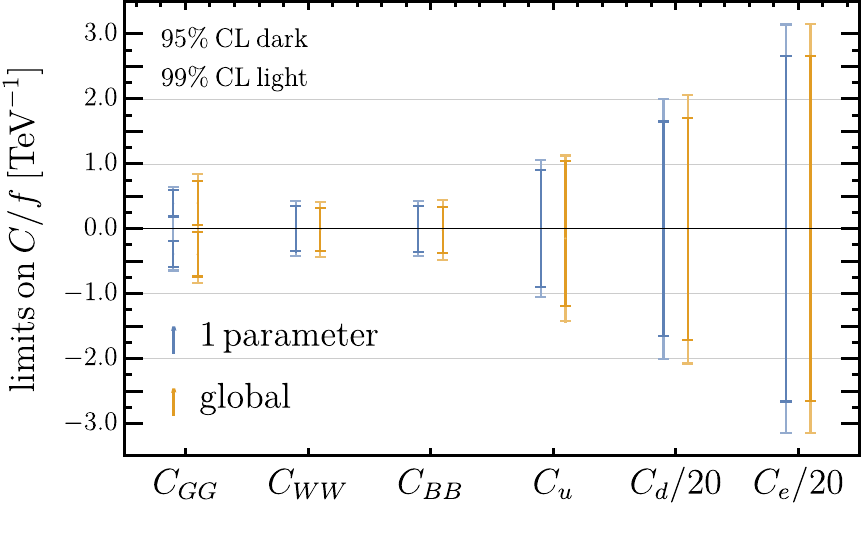}
    \caption{$95\%$ CL (dark bars) and $99\%$ CL limits (light bars) on the ALP couplings $C/f$ evaluated at the scale $\Lambda=4\pi f$ with $f=1$\,TeV. Bars in blue show the constraints obtained from individual one-parameter fits, while bars in orange refer to a global analysis marginalizing over the remaining parameters. The limits for $C_d$ and $C_e$ have been rescaled by a factor of $1/20$.}
    \label{fig:limits_indiv_vs_global}
\end{figure}
We present the limits on the ALP couplings at $95\%$~CL and $99\%$~CL in Figure~\ref{fig:limits_indiv_vs_global}. Assuming $f=1\,\text{TeV}$, we obtain $\mathcal{O}(1)$ bounds for $C_{GG}$, $C_{WW}$, $C_{BB}$ and $C_u$, while $C_d$ and $C_e$ are much less constrained, with limits of $\mathcal{O}(50)$. Our choice of the scale $f$ is motivated by the hope that new physics beyond the SM exists at a scale $\Lambda=4\pi f\approx 10$\,TeV, as motivated by the hierarchy problem in light of current LHC results.
Comparing the bounds from fitting one parameter at a time to those from a global analysis, we find that the limits on $C_{WW}$, $C_{BB}$, $C_d$, and $C_{e}$ are minimally affected by the global analysis. Importantly, the weak constraints on $C_d$ and $C_e$ do not invalidate the limits imposed on the other Wilson coefficients in the global analysis. The global bounds on $C_u$ and $C_{GG}$ are weakened by $18\%$ and $25\%$ with respect to their one-parameter counterparts. For $C_{GG}$, we notice that the corresponding limit favors a non-zero value at $95\%$~CL, but it is compatible with zero at $99\%$ CL. This discrepancy is caused by a minor experimental anomaly in three highly correlated bins of the CMS simplified template cross section analysis in the $h\to ZZ$ channel~\cite{CMS:2021ugl}, which favors non-zero values of $C_{uH}$ or $C_{uG}$ in the SMEFT and consequently shifts $C_{GG}$ away from zero. 

\begin{figure}
    \centering
    \includegraphics[width=\textwidth]{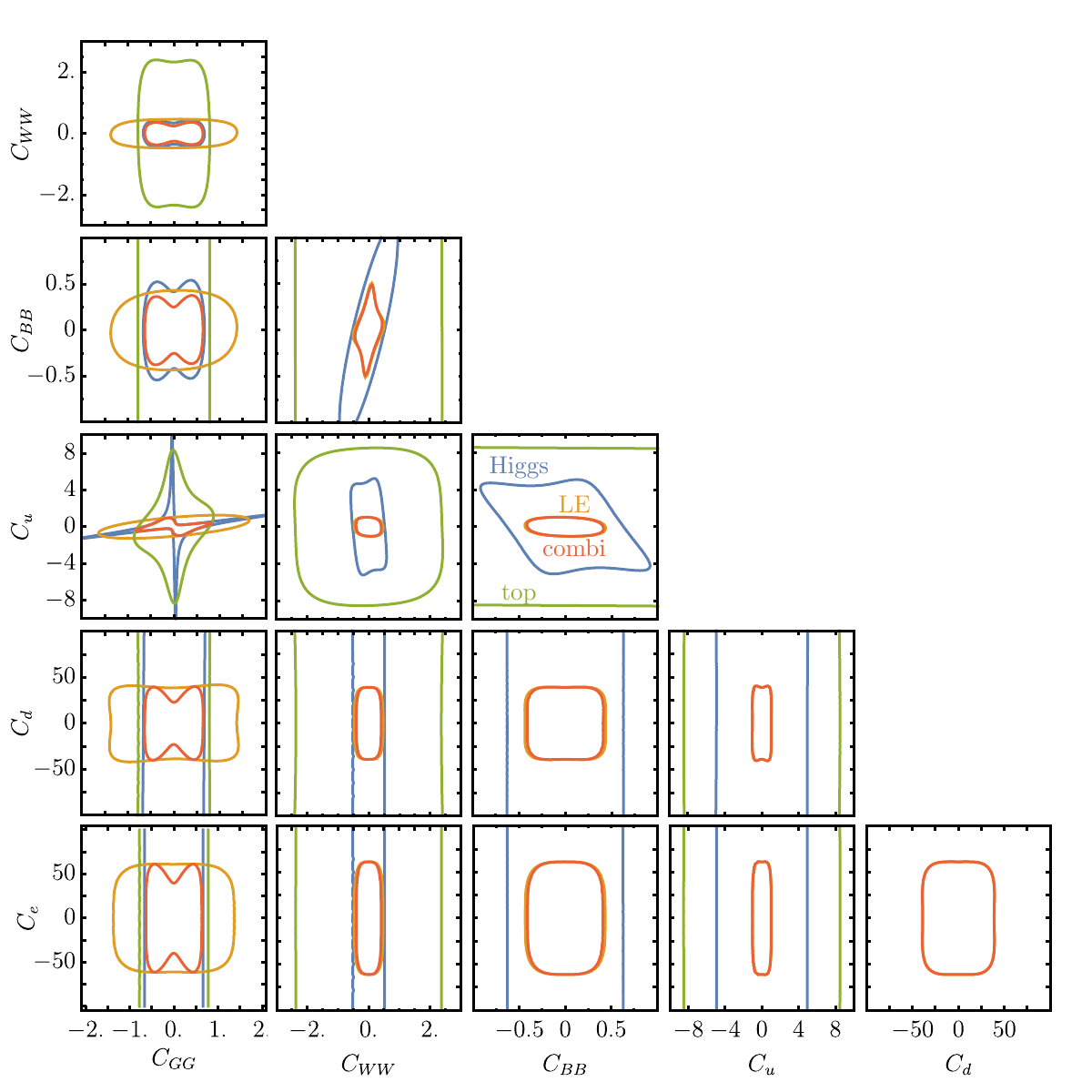}
    \caption{Two-dimensional $95\%$ CL limits on the ALP couplings evaluated at the scale $\Lambda=4\pi f$ with $f=1$\,TeV, where all other Wilson coefficients are set to zero in each panel. Different colors represent the limits from different experimental sources: low-energy (LE) data (orange), Higgs data (blue), and top data (green). The bounds derived from the combination of all experimental data are shown in red.}
    \label{fig:triangle_relevant}
\end{figure}

To investigate the correlations among the ALP couplings, we present two-dimensional fits in Figure~\ref{fig:triangle_relevant}. In each panel, we show the $95\%$~CL bounds on two Wilson coefficients while setting the remaining coefficients to zero. As expected from the mild change of the global limits in Figure~\ref{fig:limits_indiv_vs_global} with respect to the one-parameter fits, we find only weak correlations between most parameters. This is not surprising, since only few source terms of the SMEFT Wilson coefficients relevant for our analysis contain products of different ALP Wilson coefficients. Indeed, for the SMEFT Wilson coefficients appearing in our analysis, only $C_{HWB}$ as well as the dipole operators $[C_{uG}]_{33}$, $[C_{uW}]_{33}$, $[C_{uB}]_{33}$ contain the products $C_{WW}\,C_{BB}$, and $C_{u}\,C_{GG}$, $C_{u}\,C_{WW}$, $C_{u}\,C_{BB}$, respectively, in the leading-logarithmic (LL) approximation. The most interesting correlation patterns are observed for the combinations $C_{GG}$\,--\,$C_u$ and $C_{WW}$\,--\,$C_{BB}$. For $C_{GG}$\,--\,$C_u$, a free-floating $C_{GG}$ allows $C_u$ to extend into a wider parameter region. The sign of the product of $C_{GG}$ and $C_{u}$ is however constrained to be negative at $95\%$~CL when using the full data set. For $C_{WW}$ and $C_{BB}$ we find a slight preference for the product of the two coefficients to be positive. 

It is interesting to study which of the considered data sets is the driving factor in constraining the individual Wilson coefficients. Individual bounds from low-energy, Higgs and top data sets are shown in Figure~\ref{fig:triangle_relevant}. Intriguingly, low-energy data dominate the constraints on $C_{WW}$, $C_{BB}$, $C_{e}$, and even $C_u$, which to first approximation describes the ALP--top coupling. As we discuss in the next subsection, the root of the strong constraint on $C_u$ from low-energy data is that it mixes under RG evolution with $C_{HD}$, which is strongly constrained by the measurement of the $W$-boson mass. Only the bound on the ALP--gluon coupling $C_{GG}$ receives important contributions from Higgs and top data. There is an interesting interplay between the bounds from different experiments for $C_{WW}$\,--\,$C_{BB}$ and $C_{GG}$\,--\,$C_u$. For $C_{WW}$\,--\,$C_{BB}$, Higgs data allow for a relatively wide parameter range as long as their product is positive. This degeneracy for $C_{WW}$ and $C_{BB}$ is broken by low-energy data. For the pair $C_{GG}$\,--\,$C_u$, top data slightly favor a positive product of $C_{GG}$ and $C_u$, while Higgs data (as well as the combination of Higgs and top data) favor a negative product. 

\subsubsection*{Leading-log approximation vs resummation}

It is interesting to investigate how the bounds obtained from an exact solution in \eqref{eq:Evolution}, in which the leading-logarithmic corrections are resummed to all orders, differ from those derived using the LL approximation truncated at one-loop order, which leads to the simple formula
\begin{equation}
 C^\mathrm{SMEFT}_i (\mu) \approx \gamma^{\rm SMEFT-ALP}_{i\alpha\beta}(\Lambda)\,C^{\rm ALP}_\alpha(\Lambda)\,[C^{\rm ALP}_\beta(\Lambda)]^*\,\ln\frac{\mu}{\Lambda} \,.
\end{equation}
For the low-energy observables at the $Z$~pole, which dominate the fit for all coefficients except $C_{GG}$, a full list of LL parametrizations can be found in Appendix~\ref{sec:LE_Expressions}. We show in Figure~\ref{fig:limits_LL_running} the one-parameter limits on the ALP couplings obtained using the one-loop truncated LL approximation alongside with the constraints obtained from a full (resummed) evaluation of the scale evolution. For $C_{WW}$, $C_{BB}$, $C_d$, and $C_e$, the LL approximation captures the dominant effects and the limits only change marginally with respect to the resummed evolution. However, for $C_{GG}$ and $C_u$ the LL approximation lacks important effects. To investigate this further, we display the limits on $C_u$ and $C_{GG}$ from different experimental sources in the two rightmost panels of Figure~\ref{fig:limits_LL_running}. 
We observe that the limits on $C_u$ from the top and Higgs sectors remain largely unchanged when using the LL approximation for the running. The main discrepancy arises from low-energy data, where the resummed running imposes tighter constraints than in the LL approximation. This discrepancy primarily originates from the RG evolution of the Wilson coefficient $C_{HD}$. 
While the ALP contribution from $C_u$ to the Wilson coefficient $C_{HD}$ vanishes at LL order, it is simple to see that this is not the case for the resummed result. The full RG equation for $C_{HD}$ (neglecting for simplicity contributions proportional to $\alpha_i\neq\alpha_s$ and all Yukawa couplings except for $y_t$) reads~\cite{Jenkins:2013wua, Alonso:2013hga}
\begin{equation}
  \frac{d}{d\ln\mu}\,C_{HD} = \left( \frac{3\,\alpha_t}{\pi} + \frac{3\,\lambda}{8\pi^2} \right) \,C_{HD}+ \frac{6\,\alpha_t}{\pi}\,[C_{Hq}^{(1)}]_{33} - \frac{6\,\alpha_t}{\pi}\,[C_{Hu}]_{33}\,,
\end{equation}
with $\alpha_t\equiv y_t^2/(4\pi)$. The relevant ALP-induced terms that enter this RG equation are \cite{Galda:2021hbr} \begin{align}
\frac{d}{d\ln\mu}\,[C_{Hq}^{(1)}]_{33} = - \pi\,\alpha_t\,C^2_u + \cdots \,, \qquad 
\frac{d}{d\ln\mu}\,[C_{Hu}]_{33} = 2\pi\,\alpha_t\,C_{u}^2 + \cdots \,, 
\end{align}
where the ellipses refer to pure SMEFT contributions. Neglecting the running of the SM parameters results in the lowest-logarithmic approximation
\begin{align}
	C_{HD}(\mu) &= -9 \, \alpha_t^2\,C_{u}^2\,\ln^2\frac{\mu}{\Lambda} \,,
\end{align}
which is a two-loop effect. Thus, bounds on $C_{HD}$ only play a role in restricting the values of $C_u$ beyond the strict LL approximation. Even though this is a subleading effect, strong bounds on $C_{HD}$ from low-energy observables render it phenomenologically relevant for constraining $C_u$. 

For $C_{GG}$, shown in the right-most panel of Figure~\ref{fig:limits_LL_running}, the LL approximation is able to reproduce the constraints originating from the top sector quite well. Limits from low-energy data vanish completely in the LL approximation, but this data set only plays a marginal role in constraining $C_{GG}$ and hence does not influence the combined bounds. The dominant contribution to the combined constraints comes from Higgs data in this case. However, the LL approximation misses the most significant contributions from $C_{HG}$ and $[C_{uG}]_{33}$ for this data set, which are tightly constrained through gluon-fusion Higgs production and are only sourced by $C_{GG}$ beyond LL order. At lowest-logarithmic order, and taking into account the running of $C_u$ \cite{Bauer:2020jbp}, the solutions for $C_{HG}$, $[C_{uG}]_{33}$ in terms of $C_{GG}^2$ are given by 
\begin{align}
    [C_{uG}]_{33}(\mu) \supset - \frac{25 \,g_s \,y_t\, \alpha_s}{ \pi } \,C_{GG}^2\, \ln^2\frac{\mu}{\Lambda} \,, \qquad
    C_{HG}(\mu) \supset \frac{100\,\alpha_s^2\,\alpha_t}{3}\,C_{GG}^2\,\ln^3\frac{\mu}{\Lambda} \,.
\end{align}
Both of the above estimates agree very well with the resumed results from the evolution tensor $U^{(5,5)}$ in \eqref{eq:U55solu}.

Since most limits on the ALP couplings are well approximated at LL accuracy, we can deduce that they scale with the new-physics scale as $|C_i|\sim \Lambda\,\ln^{-1/2} (\Lambda/m_Z)$. As expected from the discussion above, the two exceptions to this scaling are $C_u$, for which we find $|C_u|\sim \Lambda\,\ln^{-1} (\Lambda/m_Z)$, and $C_{GG}\sim \Lambda\,\ln^n (\Lambda/m_Z)$ with $n=-3/2$ or $n=-1$, depending on whether $[C_{uG}]_{33}$ or $C_{HG}$ dominates.

\begin{figure}
    \centering
    \includegraphics[height=4.5cm]{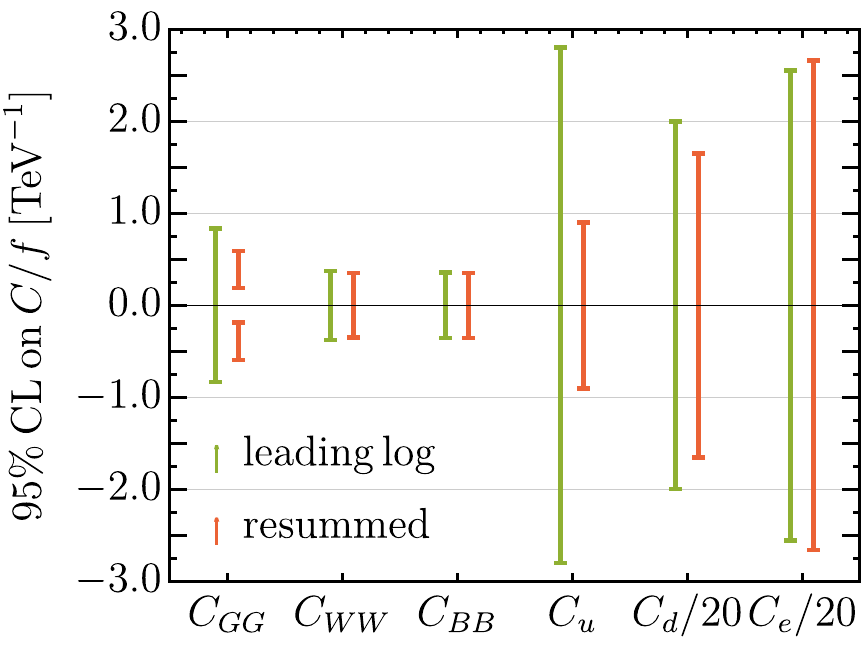}
    \quad
    \includegraphics[height=4.5cm]{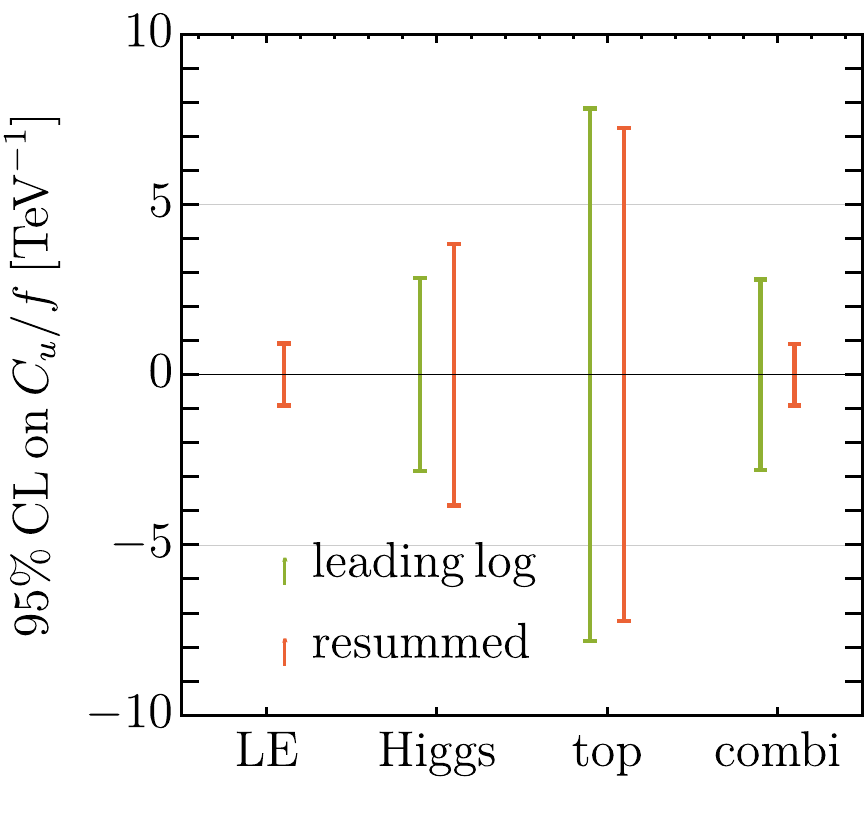}
    \quad
   \includegraphics[height=4.5cm]{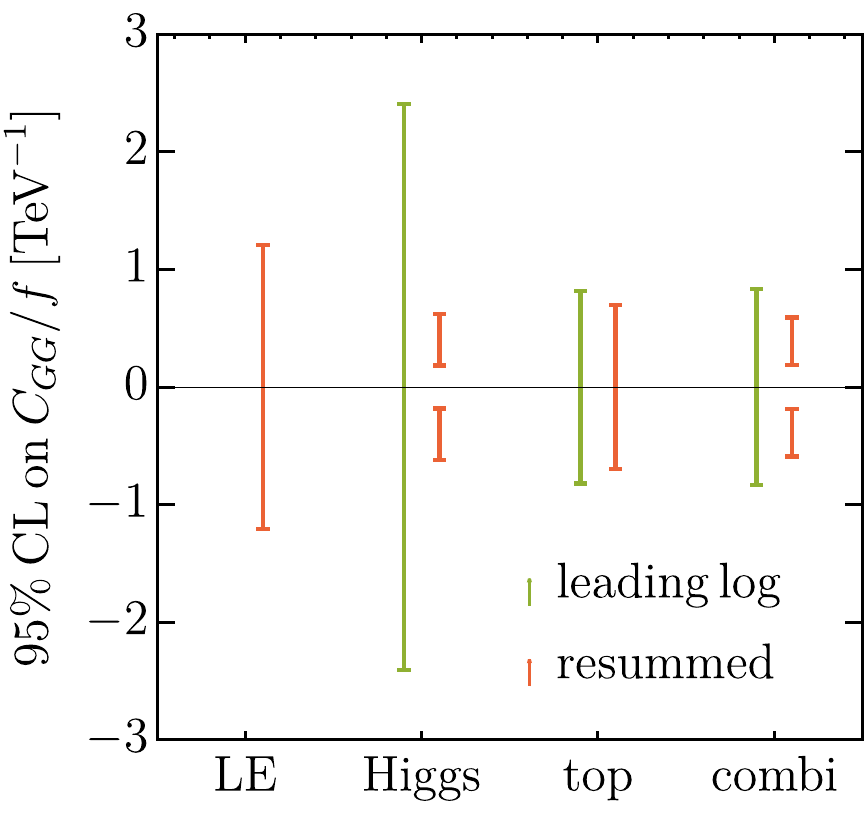}
    \caption{One-parameter fit limits on the ALP couplings using the one-loop truncated leading-logarithmic approximation (leading log) or the exact solution to the RG equations (resummed). The last two plots show the limits on $C_u$ and $C_{GG}$ derived from different experimental sources.}
    \label{fig:limits_LL_running}
\end{figure}

\subsection{Comparison with bounds from direct searches}
\label{sec:comparison_with_direct_bounds}

In this subsection, we compare our indirect limits to direct limits on ALP couplings obtained in the literature. Since direct bounds are typically stronger for light ALPs, we focus on $\mathcal{O}(\mathrm{GeV})$ ALP masses, where we expect our indirect, (mostly) mass-independent limits to be more competitive. Direct limits in this regime are dominated by collider~\cite{Bauer:2017ris} and flavor~\cite{Bauer:2021mvw} experiments.\footnote{The results in \cite{Bauer:2021mvw} are given in terms of the Wilson coefficients in \eqref{eq:lag1}. Therefore, a translation between their notation and ours is needed to compare the constraints, leading to apparent weaker limits in some cases.} 
\begin{figure}[t]
    \centering
    \includegraphics[width=.32\textwidth]{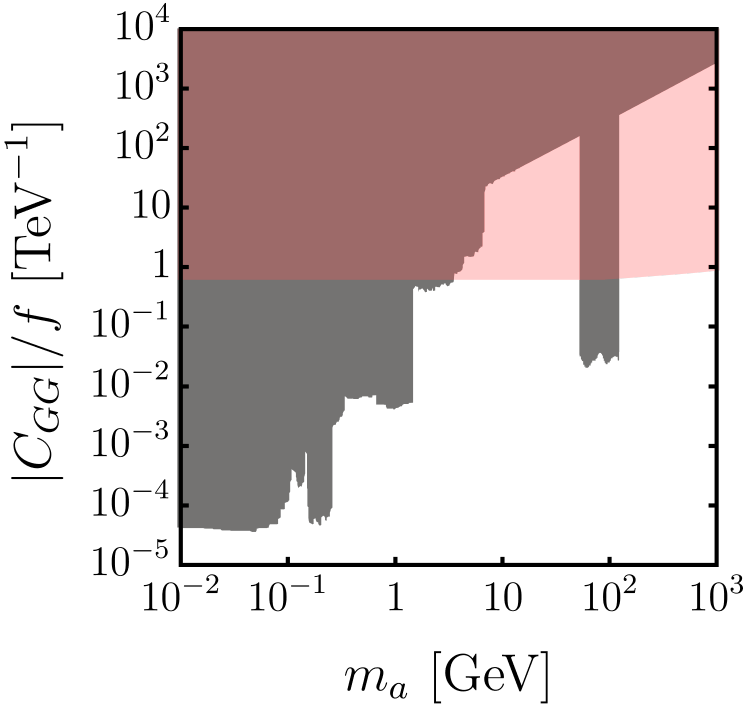}
    \includegraphics[width=.32\textwidth]{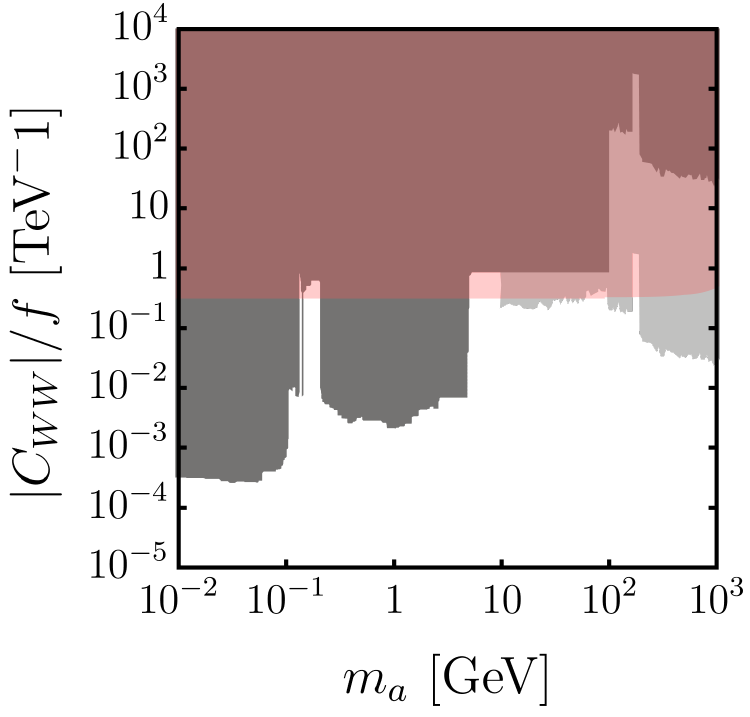}
    \includegraphics[width=.32\textwidth]{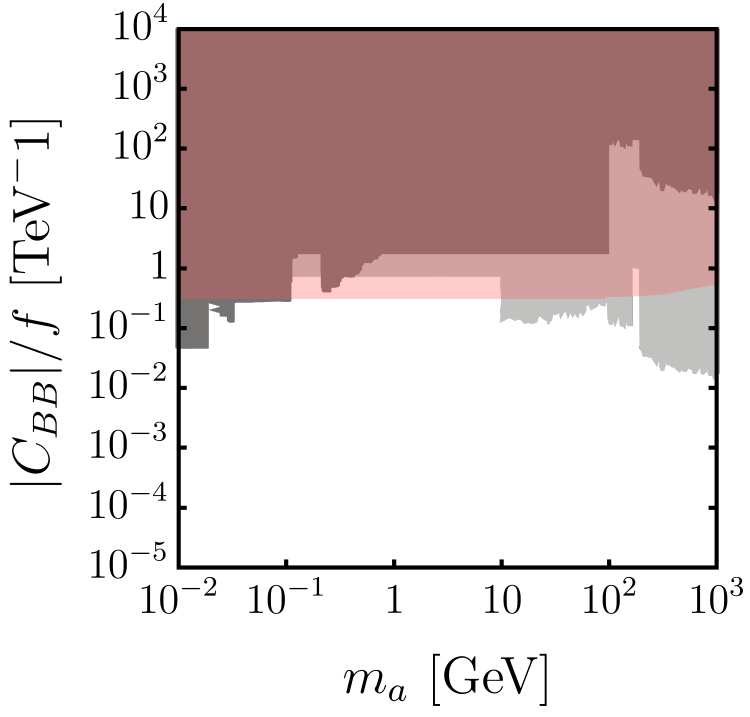}\\
    \includegraphics[width=.32\textwidth]{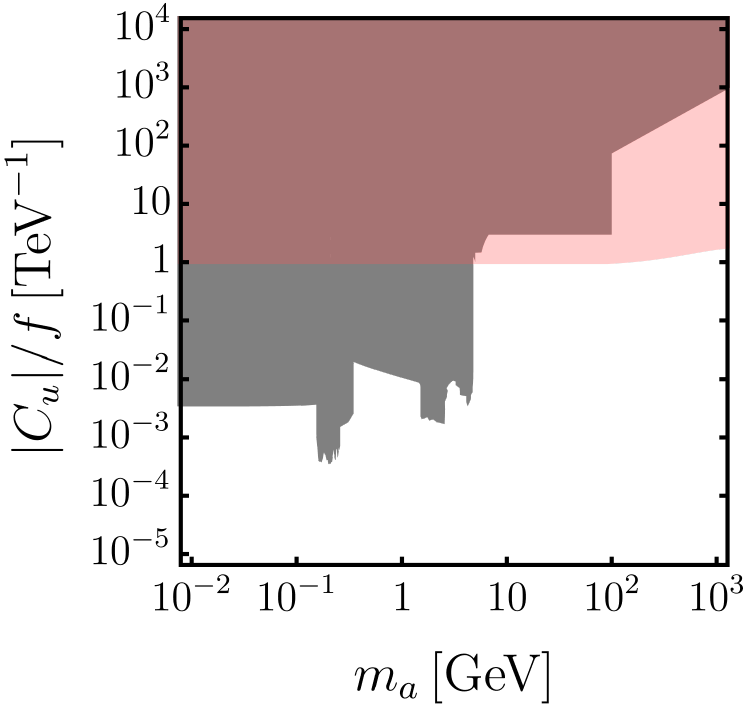}
    \includegraphics[width=.32\textwidth]{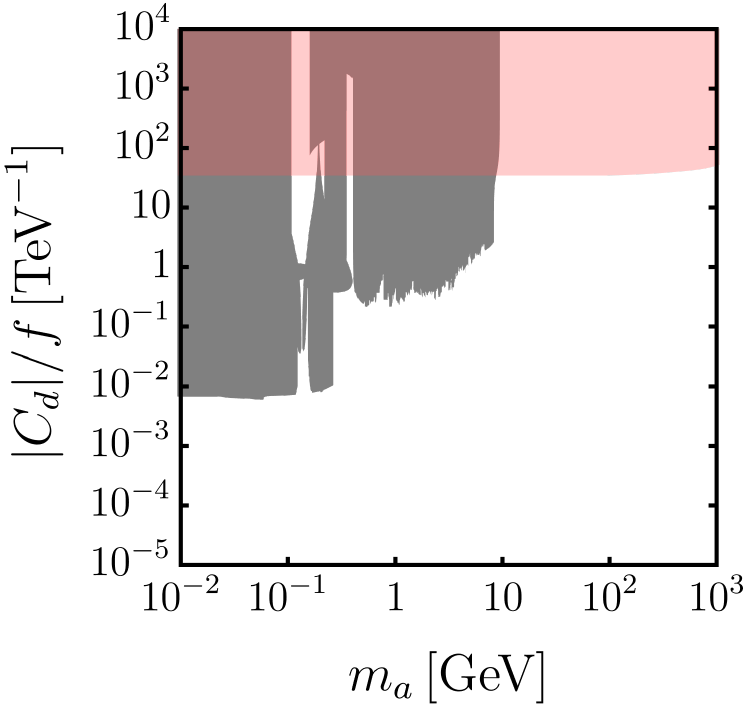}
    \includegraphics[width=.32\textwidth]{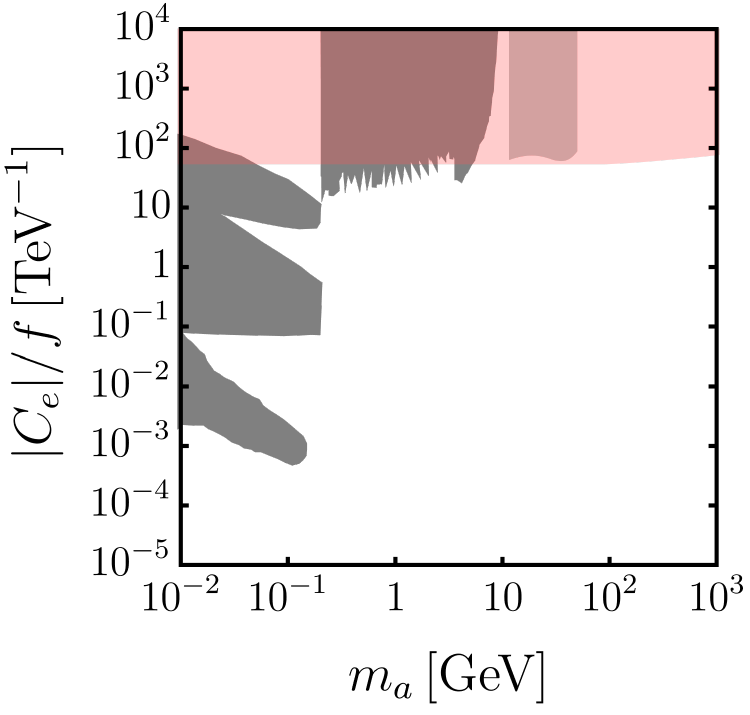}
    \caption{Indirect $95\%$ CL limits from ALP\,--\,SMEFT interference (red) compared to direct bounds from flavor, beam dump, and collider experiments, as well as supernova data. All direct bounds assume the remaining ALP Wilson coefficients to be zero. Direct bounds shaded in light gray are subject to additional model assumptions, see text for details. Note that we do not show the lower bound on $C_{GG}$, excluding a non-zero value at $95\%$ CL, as this bound disappears at $99\%$ CL. }
    \label{fig:bounds_ma_dep}
\end{figure}
We compare the direct and indirect bounds on the ALP Wilson coefficients in Figure~\ref{fig:bounds_ma_dep}, where the indirect bounds obtained from our global analysis are shown in red. The most relevant direct constraints on each of the individual ALP couplings are the following:
\begin{itemize}
    \item \textbf{$\boldsymbol{C_{GG}}$:} Flavor data impose the strongest direct constraints across the majority of the depicted ALP mass range. However, around $m_a \sim 100$ GeV, the inclusion of LHC multijet constraints becomes crucial~\cite{Mariotti:2017vtv}.
    
    \item \textbf{$\boldsymbol{C_{WW}}$ and $\boldsymbol{C_{BB}}$:} In addition to flavor constraints, non-resonant ALP contributions to vector-boson scattering yield mass-independent direct bounds on ALPs with masses up to $\sim 100$ GeV~\cite{Bonilla:2022pxu}. 
    For non-resonant gluon-fusion ALP production followed by its decay to gauge bosons~\cite{Gavela:2019cmq,Carra:2021ycg,CMS:2021xor}, constraints are placed on the products of the ALP--gluon coupling with the ALP--photon, $C_{\gamma\gamma}= s_W^2\,C_{WW} + c_W^2\,C_{BB}$, or the ALP--$Z$, $C_{ZZ}= c_W^2\,C_{WW} + s_W^2\,C_{BB}$, couplings.\footnote{Here, $c_W$ ($s_W$) is the cosine (sine) of the weak-mixing angle.} Specifically, $|C_{ZZ}\,C_{GG}|/f^2 < 4\cdot10^{-2}$~TeV$^{-2}$~\cite{CMS:2021xor} and $|C_{\gamma \gamma}\, C_{GG}|/f^2 < 5 \cdot 10^{-3}$~TeV$^{-2}$~\cite{Gavela:2019cmq}. As these bounds involve products of ALP parameters, they are not shown in our plots which depict the bounds for one coupling at a time. However, we point out that they provide the dominant bounds in the limit in which both Wilson coefficients are sizable. In the high ALP-mass region, we present bounds derived from collider constraints on the ALP--photon coupling~\cite{Bauer:2018uxu}. The corresponding limits are shown in light gray to highlight their model dependence. It is important to note that for heavy ALPs with $m_a > m_Z$, where additional decay channels like $a \to Z \gamma$ open up, a dominant decay to photons becomes highly unlikely~\cite{Alonso-Alvarez:2018irt}. In darker gray, we present the same limits assuming a branching ratio into photons of $10^{-3}$.
    
    \item \textbf{$\boldsymbol{C_u}$ and $\boldsymbol{C_d}$:} Constraints on the ALP parameter space stem from flavor data that, for the case of $C_u$, cover the full displayed ALP-mass range. In addition, $C_u$ gets further constrained by LHC $t\Bar{t}$~searches for $m_a \leq 100$~GeV~\cite{Esser:2023fdo}.

    \item \textbf{$\boldsymbol{C_e}$:} We present direct bounds from flavor and dark photon searches at BaBar~\cite{BaBar:2014zli}, which cover a similar mass range compared to constraints from $\Upsilon$ decays. Furthermore, we include constraints from SN1987A supernova observations~\cite{Lucente:2021hbp} and beam dump searches at SLAC~\cite{Essig:2010gu}, which are relevant in the $m_a < 1$~GeV mass range.     
    In the $11.5-50$~GeV mass range, we consider LHC constraints on $h \to a \mu \Bar{\mu}$~\cite{Biekotter:2022ovp}, assuming a $100\%$ branching ratio of the ALP to muons. These constraints are shaded in a lighter gray to account for the possibility of decays to taus, which would weaken the bounds.
\end{itemize}

Overall, we find that the ALP\,--\,SMEFT interference can constraint previously untested regions of the ALP parameter space. Furthermore, most direct bounds have specific model assumptions, often requiring all remaining coefficients to be zero, that do not apply to our indirect bounds. The indirect limits presented here thus offer good complementary probes, even in cases where the direct limits would a priori seem more competitive. 

\section{Interpretation in terms of UV-complete models}\label{sec:ALP_UV}

We now reinterpret our bounds in terms of UV-complete ALP models. In particular, we describe the effect of including SMEFT contributions beyond the RG-induced effects considered before, which stem from the incorporation of (tree-level) threshold corrections. We focus in the two main (fundamental) axion UV completions: the Kim-Shifman-Vainshtein-Zakharov (KSVZ)~\cite{Kim:1979if,Shifman:1979if} and the Dine-Fischler-Srednicki-Zhitnitsky (DFSZ)~\cite{Dine:1981rt,Zhitnitsky:1980tq} models, which have been already proposed as ALP benchmark scenarios~\cite{Arias-Aragon:2022iwl}.

\subsection{KSVZ model}

The KSVZ model extends the SM with a pair of fermions, $Q_L$ and $Q_R$, which transform non-trivially under $SU(3)_c$ and chirally under a global $U(1)_A$ symmetry, and a SM-singlet scalar $S$, which is only charged under the $U(1)_A$ symmetry and acquires a non-zero vacuum expectation value (vev), thus spontaneously breaking the global symmetry. The most general Lagrangian for this model reads
\begin{align}
\mathcal{L}_{\rm KSVZ}&=\mathcal{L}_{\rm SM}+|\partial_\mu S|^2+\bar Q\,i\slashed{D}\, Q- y_Q\left(S\,\bar Q_L Q_R+\mathrm{h.c.}\right)\nonumber\\
&\quad+\mu_S^2 |S|^2-\frac{\lambda_S}{2}|S|^4-\lambda_{SH}|S|^2 (H^\dagger H)+\mathcal{L}_{Qq}\,,
\end{align}
where $y_Q$, $\mu_S$, $\lambda_S$, and $\lambda_{SH}$ are real parameters and $\mathcal{L}_{Qq}$ is a possible portal coupling between $Q$ and a SM fermion. As discussed in~\cite{DiLuzio:2017pfr}, this term is introduced to let the extra quarks decay, as otherwise the Lagrangian would be invariant under a vectorial $U(1)_Q$ symmetry that would make them stable. In the original KSVZ implementation, where the extra quarks transform under the SM as $Q_{L,R}\sim(\mathbf{3},\mathbf{1})_0$, no renormalizable term for $\mathcal{L}_{Qq}$ is possible. However, there are multiple representation choices for which $\mathcal{L}_{Qq}\neq 0$. For concreteness, we consider here the case where $Q_{L,R}\sim(\mathbf{3},\mathbf{1})_{-1/3}$ and the global $U(1)_A$ charges are $X_S=1$, $X_{Q_L}=1$, and $X_{Q_R}=0$. In this case, we have
\begin{align}
\mathcal{L}_{Qq}=-y_q^p\, \bar q_L^p H Q_R +\mathrm{h.c.} \,.
\end{align}
Additionally, we consider possible soft $U(1)_A$-breaking terms that will give mass to the pseudo-Goldstone ALP at the expense of spoiling the solution to the strong CP problem. For definiteness, we consider the term
\begin{align}
\mathcal{L}=\mathcal{L}_{\rm KSVZ}+\frac{\kappa^2}{2}(S^2+S^{*\,2})\,,
\end{align}
with $\kappa$ being a real parameter controlling the size of the breaking.

In the $U(1)_A$-broken phase, it is convenient to write the scalar singlet as 
\begin{align}
S(x)=\frac{1}{\sqrt{2}}\big[f+\rho(x)\big]\,e^{\frac{ia(x)}{f}}\,,
\end{align}
with $f$ denoting the vev of $S$, and $\rho$ and $a$ corresponding to the radial and pseudo-Goldstone components of the field, respectively. After performing the fermion shift
\begin{align}
Q_L\to e^{\frac{ia}{f}}\,Q_L\,,
\end{align}
which removes the ALP from the Yukawa interaction, the UV Lagrangian in the $U(1)_A$-broken phase reads
\begin{align}
\mathcal{L}&=\mathcal{L}_{\rm SM}-\frac{\lambda_{SH}f^2}{2}(H^\dagger H)+\frac{1}{2}(\partial_\mu\rho)^2+\frac{1}{2}\Big(1+\frac{\rho}{f}\Big)^2(\partial_\mu a)^2+\bar Q\,i\slashed{D}\, Q-\frac{\partial_\mu a}{f}\,\bar Q_L\gamma^\mu Q_L\nonumber\\
&\quad -\frac{a}{f}\, \frac{\alpha_s}{8\pi}\, G_{\mu\nu}^A \widetilde G^{\mu\nu\,A}-\frac{1}{3}\,\frac{a}{f}\, \frac{\alpha_Y}{4\pi}\, B_{\mu\nu} \widetilde B^{\mu\nu}-\left[M_Q\Big(1+\frac{\rho}{f}\Big)\bar Q_L Q_R+y_q^p\, \bar q_L^p H Q_R+\mathrm{h.c.}\right]\nonumber\\
&\quad-\frac{1}{2}M_\rho^2\,\rho^2-3\lambda_S f\,\frac{\rho^3}{3!}-3\lambda_S\,\frac{\rho^4}{4!}-\lambda_{SH}\left(f\,\rho+\frac{\rho^2}{2}\right)(H^\dagger H)-\frac{m_a^2}{4}(f+\rho)^2\,\left(1-\cos\frac{2a}{f}\right) ,
\end{align}
with $M_Q=y_Q\,f/\sqrt{2}$, $M_\rho^2=\lambda_S f^2$ and $m_a^2=2\kappa^2$. Note that, after $U(1)_A$-symmetry breaking, the Higgs doublet mass gets a correction of order $\lambda_{SH} f$. Thus, if the hierarchy between the electroweak scale and $f$ is large, one would expect $\lambda_{SH}\sim v^2/f^2$ to avoid a fine-tuned cancellation of parameters.

We consider the scenario in which $M_\rho,M_Q\sim f$ are heavy and integrate out the corresponding particles. The resulting EFT Lagrangian at tree-level order and up to dimension-six interactions reads\footnote{We have used the \texttt{Mathematica} package \texttt{Matchete}~\cite{Fuentes-Martin:2022jrf} to cross-check this matching result.}
\begin{align}
\label{eq:KSVZ_EFT}
\mathcal{L}_{\rm EFT}&=\mathcal{L}_{\rm SM}-\frac{\lambda_{SH}f^2}{2}(H^\dagger H)+\frac{1}{2}\frac{f^2\,\lambda_{SH}^2}{M_\rho^2}(H^\dagger H)^2+\frac{1}{2}(\partial_\mu a)^2-\frac{1}{2}m_a^2\,a^2-\frac{a}{f}\, \frac{\alpha_s}{8\pi}\, G_{\mu\nu}^A \widetilde G^{\mu\nu\,A}\nonumber\\
&\quad-\frac{1}{3}\,\frac{a}{f}\, \frac{\alpha_Y}{4\pi}\, B_{\mu\nu} \widetilde B^{\mu\nu}+4\,\frac{m_a^2}{f^2}\,\frac{a^4}{4!}+\lambda_{SH}\,\frac{m_a^2}{M_\rho^2}\,a^2(H^\dagger H)-\frac{\lambda_{SH}}{M_\rho^2}\,(\partial_\mu a)^2 (H^\dagger H)\nonumber\\
&\quad-\frac{\lambda_{SH}^2f^2}{2M_\rho^4}Q_{H\Box}+\frac{y_q^{p}y_q^{r\,*}}{2M_Q^2}\left(\bm{Y}_d^{rs}\, [Q_{dH}]^{ps}-\frac{1}{2}\,[Q_{Hq}^{(1)}]^{pr}-\frac{1}{2}\,[Q_{Hq}^{(3)}]^{pr}+\mathrm{h.c.}\right) ,
\end{align}
where the second and third terms are removed by an appropriate redefinition of the Higgs potential parameters. Namely,
\begin{align}
\mu^2\to\tilde\mu^2=\mu^2-\frac{\lambda_{SH}f^2}{2}\,, \qquad
\lambda\to\tilde\lambda=\lambda-\frac{f^2\,\lambda_{SH}^2}{2 M_\rho^2}\,.
\end{align}
Furthermore, we see that the explicit $U(1)_A$-breaking term not only gives mass to the ALP, but also introduces other shift-symmetry breaking interactions, $m_a^2\,a^4$ and $m_a^2\,a^2(H^\dagger H)$.\footnote{These additional shift-symmetry breaking interactions do not alter our results in Sections~\ref{sec:ALPCouplings} and~\ref{sec:global_analysis}, except for the running of $m_H$ in \eqref{eq:SM_ALP_running}, which receives similar effects to those from $c_{HH}$.} Finally, $Q_i$ are dimension-six SMEFT operators whose definition can be found in~\cite{Grzadkowski:2010es}.

We now turn to analyzing constraints on the KSVZ model from Higgs, top and low-energy data. As discussed above, the KSVZ model features a fermiophobic axion (at tree-level order), with different $Q$ charges under the SM gauge group yielding different values $C_{GG}$, $C_{WW}$ and $C_{BB}$. A common feature of all KSVZ models is the presence of a non-zero $Q_{H\square}$, which is generated via the tree-level exchange of the associated scalar radial excitation. Profiling over the remaining parameters, we obtain the constraint $\lambda_S^2\, f/\lambda_{SH} > 2.8 \,\text{TeV}$ from the limits on the $Q_{H\square}$ coefficient. On the other hand, the limits on the bosonic ALP couplings when profiling over the remaining parameters (including the Wilson coefficient of $Q_{H\square}$) do not change by more than $10\%$ with respect to the one-parameter limits presented in Figure~\ref{fig:limits_indiv_vs_global}. Therefore, we refer to this plot for limits on KSVZ models with generic $Q$ charges. 

For KSVZ models with additional portal couplings for the heavy vector-like quarks, such as the one presented in \eqref{eq:KSVZ_EFT}, we can additionally set constraints on the coupling strength of the portal coupling $y_q$. Assuming for simplicity that $y_q$ is flavor universal, we find the limit $|y_q/M_Q| < 0.1$~TeV$^{-1}$, which is dominated by the strong constraints on $C_{Hq}^{(1)}$ and $C_{Hq}^{(3)}$ from electroweak precision observables.

\subsection{DFSZ model}

The DFSZ models consists of a two-Higgs-doublet, $H_{1,2}$, plus SM-singlet, $S$, scalar extension of the SM. The Lagrangian of the model is chosen such that it preserves, at the classical level, a global $U(1)_A$ symmetry and reads
\begin{align}
\mathcal{L}_{\rm DFSZ}&\supset |D_\mu H_1|^2+|D_\mu H_2|^2+|\partial_\mu S|^2-(\bar q\,\tilde H_1 \,\bm{\Gamma}_u\,u_R+\bar q\,H_2\,\bm{\Gamma}_d\, d_R+\bar \ell\,H_i\,\bm{\Gamma}_e\, e_R+\mathrm{h.c.})\nonumber\\
&\quad-m_1^2\,|H_1|^2-m_2^2\,|H_2|^2-\frac{\lambda_1}{2}\,|H_1|^4-\frac{\lambda_2}{2}\,|H_2|^4-\lambda_3\,|H_1|^2|H_2|^2-\lambda_4\,|H_1^\dagger H_2|^2\nonumber\\
&\quad+\mu_S^2\,|S|^2-\frac{\lambda_S}{2}\,|S|^4-\lambda_{SH_1}\,|S|^2 |H_1|^2-\lambda_{SH_2}\,|S|^2 |H_2|^2-\lambda_{SH_{12}}\left[(H_1^\dagger H_2)S^2+\mathrm{h.c.}\right],
\end{align}
where $\supset$ denotes that we omitted the SM-like terms in the Lagrangian. All scalar potential parameters in the Lagrangian above are real, including $\lambda_{SH_{12}}$ which can be made real by appropriate global phase redefinitions of the fields. In the charged-lepton Yukawa, $i=1,2$ corresponds to two different versions of the model, which we denote as DFSZ I and II, respectively. The last term also admits a different choice, with $S$ rather than $S^2$, corresponding to a different $U(1)_A$ charge implementation. Different choices for this term have mild effects in the ensuing discussion and we focus on this variant of the model for definiteness. As we did for the KSVZ model, we admit the possibility of an explicit $U(1)_A$-breaking term, which we choose to be the same as before
\begin{align}
\mathcal{L}=\mathcal{L}_{\rm DFSZ}+\frac{\kappa^2}{2}(S^2+S^{*\,2})\,,
\end{align}
with $\kappa$ being a real parameter controlling the size of the breaking.

As before, the scalar potential parameters are chosen such that $S$ acquires a vev that breaks the global $U(1)_A$ symmetry. Once more, we parameterize the SM-singlet as
\begin{align}
S(x)=\frac{1}{\sqrt{2}}\,\big[f+\rho(x)\big]\,e^{\frac{ia(x)}{f}}\,.
\end{align}
The two-Higgs-doublet spectrum can be rather different depending on the values of $m_1$, $m_2$ and $\lambda_{SH_i}$. Here, we assume that these parameters are such that we are in a decoupling regime where a full doublet and $\rho$ are much heavier than the ALP and the SM particles.\footnote{It would be interesting to consider a low-scale two-Higgs-doublet regime, see e.g.~\cite{Choi:2017gpf,Alonso-Alvarez:2021ett}, where we depart from our original assumption that the ALP\,--\,SMEFT Lagrangian in~\eqref{eq:ALP_lag} is the relevant EFT. This would require extending the present EFT framework and is beyond the scope of this work.} The heavy doublet, $\Phi$, and SM Higgs, $H$, are linear combinations of $H_1$ and $H_2$. Namely,
\begin{align}\label{eq:HiggsRot}
\begin{pmatrix}
H_1\\
H_2
\end{pmatrix}
=
R(\alpha)
\begin{pmatrix}
H\\
\Phi
\end{pmatrix}
\quad\textrm{such that}\quad
R(\alpha)^T
\begin{pmatrix}
m_{11}^2 & m_{12}^2\\[5pt]
m_{12}^2 & m_{22}^2
\end{pmatrix}
R(\alpha)=
\begin{pmatrix}
-\mu^2 & 0\\
0 & M_\Phi^2
\end{pmatrix}
,
\end{align}
where $m_{ii}^2=m_i^2+\lambda_{SH_i}\,f^2/2$ ($i=1,2$), $m_{12}^2=\lambda_{SH_{12}}f^2/2$, and $R(\alpha)$ is a $2\times2$ rotation matrix. Once this rotation is taken into account, the SM Yukawas are related to the mixing angle and the original Yukawas as
\begin{align}
\bm{Y}_u=c_\alpha\,\bm{\Gamma}_u\,,\qquad
\bm{Y}_d=s_\alpha\,\bm{\Gamma}_d\,,\qquad
\bm{Y}_e=\left\{\begin{matrix}c_\alpha\,\bm{\Gamma}_e&\textrm{DFSZ I}\\[2pt] s_\alpha\,\bm{\Gamma}_e&\textrm{DFSZ II}\end{matrix}\right.\,,
\end{align}
with $c_\alpha\equiv\cos\alpha$, and $s_\alpha\equiv\sin\alpha$. Perturbativity constraints on the UV Yukawa couplings restrict the possible values of $\alpha$. Using the perturbative unitarity bound $|\Gamma_u^{33}|\lesssim3$~\cite{Allwicher:2021rtd}, we get the constraints $|c_\alpha|\gtrsim y_t/3$ and $|s_\alpha|\gtrsim y_b/3$, independently of the DFSZ type.

The pseudo-Goldstone, $a$, can be moved away from the scalar potential, yielding a coupling structure like the one in~\eqref{eq:lag1}, by means of the following shifts of the scalar and fermion fields
\begin{align}
H_1&\to e^{\frac{ia}{f}X_{H_1}}\,H_1\,,&
H_2&\to e^{\frac{ia}{f}X_{H_2}}\,H_2\,,\nonumber\\
u_R&\to e^{\frac{ia}{f}X_{H_1}}\,u_R\,,&
d_R&\to e^{-\frac{ia}{f}X_{H_2}}\,d_R\,,&
e_R&\to e^{-\frac{ia}{f}X_{H_i}}\,e_R\,,
\end{align}
with $X_{H_1}=2s_\alpha^2$ and $X_{H_2}=-2c_\alpha^2$.\footnote{This choice of $H_{1,2}$ shifts is uniquely determined by the requirement of having no contributions to the $\partial^\mu a\,( H^\dagger i\overleftrightarrow{D_\mu} H)$ operator, which would introduce a mixing between the ALP and the $Z$ would-be Goldstone boson after electroweak symmetry breaking.} After performing these shifts and the Higgs rotation in~\eqref{eq:HiggsRot}, we obtain the following UV Lagrangian in the $U(1)_A$-broken phase
\begin{align}
\mathcal{L}&\supset\mathcal{L}_{\rm SM}+\bigg[\frac{1}{2}\Big(1+\frac{\rho}{f}\Big)^2+s_{2\alpha}^2\frac{|H|^2}{f^2}\bigg](\partial_\mu a)^2+c_u\frac{\partial_\mu a}{f}\,\bar u_R\gamma^\mu u_R+c_d\frac{\partial_\mu a}{f}\,\bar d_R\gamma^\mu d_R+c_e\frac{\partial_\mu a}{f}\,\bar e_R\gamma^\mu e_R\nonumber\\
&\quad+3\,\frac{a}{f}\, \frac{\alpha_s}{4\pi}\, G_{\mu\nu}^A \widetilde G^{\mu\nu\,A}+c_{BB}\,\frac{a}{f}\, \frac{\alpha_Y}{4\pi}\, B_{\mu\nu} \widetilde B^{\mu\nu}+|D_\mu \Phi|^2-M_\Phi\,|\Phi|^2+\frac{1}{2}(\partial_\mu\rho)^2-\frac{1}{2}M_\rho^2\,\rho^2\nonumber\\
&\quad-\left[-t_\alpha\,\bar q\,\tilde \Phi\,\bm{Y}_u\, u_R+ t_\alpha^{-1}\bar q\,\Phi \,\bm{Y}_d\,d_R+\eta_\alpha\,\bar \ell\,\Phi\,\bm{Y}_e\, e_R+\lambda_{\Phi H}\,|H|^2 (H^\dagger\Phi)+\lambda_{S\Phi H}f\,\rho(H^\dagger\Phi)+\mathrm{h.c.}\right]\nonumber\\
&\quad-3\lambda_S f\,\frac{\rho^3}{3!}-3\lambda_S\frac{\rho^4}{4!}-\frac{m_a^2}{4}(f+\rho)^2\left(1-\cos\frac{2a}{f}\right)-\lambda_{SH}\left(f\rho+\frac{\rho^2}{2}\right)|H|^2,
\end{align}
where $M_\rho^2=\lambda_S f^2$, $m_a^2=2\kappa^2$, $t_\alpha\equiv\tan\alpha$ and $\eta_\alpha=-t_\alpha$ ($t_\alpha^{-1}$) for DFSZ I (II), and we omitted the Lagrangian terms involving heavy fields (either $\Phi$ or $\rho$) that do not contribute to the tree-level matching at dimension six. The dimension-five ALP couplings depend on the model variant and are given by $c_u=-c_e=-2s_\alpha^2$, $c_d=-2c_\alpha^2$, $c_{BB}=2$ in DFSZ I, while $c_u=-2s_\alpha^2$, $c_d=c_e=-2c_\alpha^2$, $c_{BB}=8$ in DFSZ II. Finally, we have defined the following couplings for simplicity
\begin{align}
\lambda_{\Phi H}&=s_\alpha c_\alpha \left[-c_\alpha^2\,\lambda_1+s_\alpha^2\,\lambda_2+c_{2 \alpha}\,(\lambda_3+\lambda_4)\right]\,,&
\lambda_{SH}&=s_{2\alpha}\,\lambda_{SH_{12}}+c_\alpha^2\,\lambda_{SH_1}+s_\alpha^2\,\lambda_{SH_2}\,,\nonumber\\
\lambda_{S\Phi H}&=\frac{1}{2}\left[2c_{2\alpha}\,\lambda_{SH_{12}}-s_{2\alpha}\,(\lambda_{SH_1}-\lambda_{SH_2})\right]\,.
\end{align}

We integrate out the $\rho$ and $\Phi$ fields at tree-level with the help of \texttt{Matchete}~\cite{Fuentes-Martin:2022jrf}. The resulting EFT Lagrangian at dimension-six reads
\begin{align}
\label{eq:DFSZ_EFT}
\mathcal{L}_{\rm EFT}&=\mathcal{L}_{\rm SM}+\frac{1}{2}(\partial_\mu a)^2-\frac{1}{2}m_a^2\,a^2+3\,\frac{a}{f}\, \frac{\alpha_s}{4\pi}\, G_{\mu\nu}^A \widetilde G^{\mu\nu\,A}+c_{BB}\,\frac{a}{f}\, \frac{\alpha_Y}{4\pi}\, B_{\mu\nu} \widetilde B^{\mu\nu}+c_u\frac{\partial_\mu a}{f}\,\bar u_R\gamma^\mu u_R\nonumber\\
&\quad+c_d\frac{\partial_\mu a}{f}\,\bar d_R\gamma^\mu d_R+c_e\frac{\partial_\mu a}{f}\,\bar e_R\gamma^\mu e_R+4\,\frac{m_a^2}{f^2}\,\frac{a^4}{4!}-\left(\frac{\lambda_{SH}}{M_\rho^2}-\frac{s_{2\alpha}^2}{f^2}\right)(\partial_\mu a)^2 |H|^2\nonumber\\
&\quad+\lambda_{SH}\frac{m_a^2}{M_\rho^2}\,a^2\,|H|^2-\frac{C_{\psi H}}{M_\Phi^2}\left(t_\alpha\,[\bm{Y}_u]^{pr}\, [Q_{uH}]^{pr}- t_\alpha^{-1}\,[\bm{Y}_d]^{pr}\, [Q_{dH}]^{pr}-\eta_\alpha\,[\bm{Y}_e]^{pr}\, [Q_{eH}]^{pr}+\mathrm{h.c.}\right)\nonumber\\
&\quad-\frac{[\bm{Y}_u^*]^{sr}\, [\bm{Y}_u]^{pt}\,t_\alpha^2}{M_\Phi^2} \left(\frac{1}{6}\,[Q_{qu}^{(1)}]^{prst}+[Q_{qu}^{(8)}]^{prst}\right)-\frac{[\bm{Y}_d^*]^{sr}\, [\bm{Y}_d]^{pt}\,t_\alpha^{-2}}{M_\Phi^2} \left(\frac{1}{6}\,[Q_{qd}^{(1)}]^{prst}+[Q_{qd}^{(8)}]^{prst}\right)\nonumber\\
&\quad-\frac{[\bm{Y}_e^*]^{sr}\, [\bm{Y}_e]^{pt}\,\eta_\alpha^2}{2M_\Phi^2}\,[Q_{le}]^{prst}-\frac{1}{M_\Phi^2}\,\Big([\bm{Y}_u]^{pr}\, [\bm{Y}_d]^{st}\, [Q_{quqd}^{(1)}]^{prst}-[\bm{Y}_u]^{st}\, [\bm{Y}_e]^{pr}\,t_\alpha \eta_\alpha\, [Q_{lequ}^{(1)}]^{prst}\nonumber\\
&\quad-[\bm{Y}_d^*]^{st}\, [\bm{Y}_e]^{pr}\,t_\alpha^{-1} \eta_\alpha\, [Q_{ledq}]^{prst}+\mathrm{h.c.}\Big)+\frac{C_H}{M_\Phi^2}\,Q_H-\frac{\lambda_{SH}^2f^2}{2M_\rho^4}\,Q_{H\Box}\,,
\end{align}
where $C_{\psi H}$ and $C_H$ are given in terms of the original scalar-potential parameters by
\begin{align}
C_H&=C_{\psi H}^2\,,&
C_{\psi H}&=\lambda_{\Phi H}-\lambda_{SH}\,\lambda_{S\Phi H}\,\frac{f^2}{M_\rho^2}\,.
\end{align}
Analogously to what we did in the KSVZ case, we have redefined the SM Higgs potential parameters, $\mu$ and $\lambda$, to account for the matching corrections. If the hierarchy between the electroweak scale and $f$ is large, one would again expect $\lambda_{SH}\sim v^2/f^2$ to avoid a fine-tuned cancellation of scalar-potential parameters.

\begin{figure}
    \centering
    \includegraphics[height=6cm]{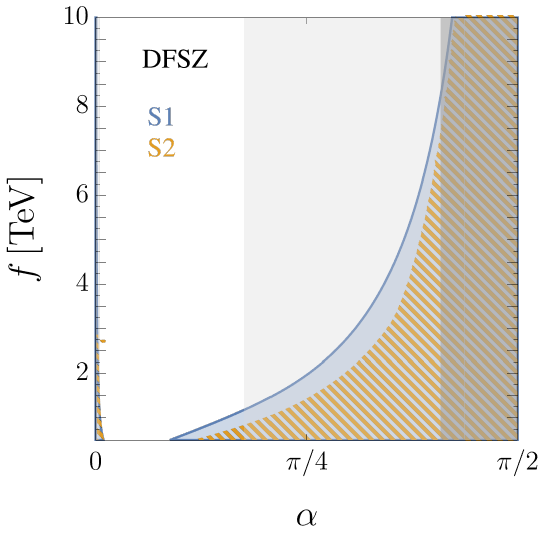}
    \caption{Two-dimensional limits on $\alpha$ vs $f$ in the DFSZ model for two benchmark scenarios: S1 (benchmark) and S2 (profiled), see text for details. The dark and light gray bands corresponds to regions where pertubative unitarity is violated (namely, $\Gamma_u^{33}\gtrsim 3$) and where $\Gamma_u^{33}\gtrsim 1$, respectively.}
    \label{fig:DFSZ_limits}
\end{figure}

Translating the above notation to the one used in our global analysis (cf.\ \eqref{eq:lag2}), we find that only ALP couplings to fermions are non-zero in both DFSZ models: $C_u=-C_e=-2s_\alpha^2$ and $C_d=-2c_\alpha^2$ for DFSZ~I, and $C_u=-2s_\alpha^2$ and $C_d=C_e=-2c_\alpha^2$ for DSFZ~II. We present the bounds on the mixing angle $\alpha$ and the ALP decay constant $f$ in Figure~\ref{fig:DFSZ_limits}. We consider two scenarios: S1) where the coefficients of the SMEFT operators $Q_H$, $Q_{H\Box}$, and $Q_{\psi H}$ are assumed to be suppressed, as would be expected if the scalar-potential parameters are small; and S2) where we profile over the coefficients of these operators within the range $|C_{\psi H}|, \, |C_{H\square}|, |C_H| <1 $. Since the ALP couplings to fermions in the DFSZ~I and DFSZ~II models differ only by their couplings to leptons, which are weakly constrained, we find that the limits on both models are (almost) identical. 
As shown in Figure~\ref{fig:DFSZ_limits}, the overall limit on $f$ is dominated by the matching correction from the four-quark operators $Q_{qu}^{(8)}$ and in particular $Q_{qu}^{(1)}$, which run into the coefficients of the SMEFT operators $Q_{HD}$, $Q_{Hq}^{(1)}$ and $Q_{Hq}^{(3)}$ that are tightly constrained at the electroweak scale. The obtained limits are found to be weak, except in the region where the UV Yukawa couplings are larger than one and dominates close to the non-perturbative Yukawa region. Limits from the ALP coupling to up-type quarks are suppressed by $s_\alpha^2$ and thus only play a subdominant role. When profiling over the other matching corrections within the range $|C_{\psi H}|, \, |C_{H\square}|, |C_H| <1 $, we observe that the limits on both DFSZ models are slightly diminished with respect to S1, especially for intermediate values of $|\alpha|$.

\section{Conclusions}
\label{sec:conclusion}

While the SMEFT is normally used to describe the possible effects of yet undiscovered heavy particles, we have shown in this paper that SMEFT analyses can also be reinterpreted to infer indirect information on light new physics. Exploiting the non-trivial RG flow of the ALP couplings into the SMEFT Wilson coefficients~\cite{Galda:2021hbr}, we have used existing SMEFT studies including low-energy, Higgs and top data to constrain these couplings. Contrary to other ALP constraints, the ones presented here posses the unique feature of being largely independent of particular assumptions on the ALP mass, lifetime or branching rations.

Furthermore, we have obtained for the first time, a semi-analytic solution to the ALP RG equations at dimension six under the assumption of flavor-universal ALP interactions. This solution, given in the form of RG evolution tensors, can readily be used to derive ALP and SMEFT couplings at an arbitrary scale (provided there are no mass thresholds). Its generalization to generic ALP interactions and the incorporation of threshold corrections into a modified version of \texttt{DsixTools} will be presented in a forthcoming publication, thus paving the way for automated ALP analyses. Even with the present assumptions, the ALP-to-SMEFT evolution tensor provided in the ancillary files can readily be used in most SMEFT analyses, such as the one presented in this paper. 

The resulting bounds on the bosonic ALP interactions $C_{GG}$, $C_{WW}$, $C_{BB}$ and the ALP coupling to up-type quarks $C_u$ are found to be of $\mathcal{O}(1)$ for $f=1$\,TeV. The couplings to down-type quarks and leptons, $C_d$ and $C_e$, remain weakly constrained, with limits of $\mathcal{O}(50)$, as expected given the additional Yukawa suppression present in these couplings. The bounds on $C_{WW}$, $C_{BB}$, $C_u$, $C_d$, and $C_e$ primarily arise from low-energy precision observables, such as measurements at the $Z$~pole. On the contrary, the limit on $C_{GG}$ is mainly driven by Higgs and top physics. In our global analysis, we found weak correlations between Wilson coefficients. However, there is an interesting interplay between the limits on $C_u$ and $C_{GG}$, where a negative product of the two Wilson coefficients is favored, and on $C_{WW}$ and $C_{BB}$, where any large product of the two coefficients is ruled out. We have also found that the LL approximation captures most of the phenomenologically-relevant effects for all ALP couplings except $C_u$ and $C_{GG}$, which generate non-trivial contributions to strongly constrained SMEFT directions at higher order in the RG resummation. 

Comparing with direct ALP searches, our limits constrain large regions of previously uncovered areas of the parameter space for ALP masses above $10$~GeV. Even for lower masses, the obtained bounds can partly compete with existing ones and have the major advantage of being independent of specific assumptions on the ALP properties, thus offering a complementary probe in regions where direct bounds would a priori dominate. It would be interesting to investigate how the combination of direct and indirect bounds further narrows the ALP parameter space in a global analysis. We leave the comprehensive study of both direct and indirect constraints in a global analysis for future work. 

We have also reinterpreted our results in the context of two benchmark UV completions based on the KSVZ and DSFZ models. The KSVZ model features no tree-level couplings to fermions and we have found that tree-level threshold corrections, arising from integrating out additional heavy particles present in the model, do not significantly affect the limits on $C_{GG}$, $C_{WW}$ and $C_{BB}$ obtained from the ALP\,--\,SMEFT analysis. On the contrary, the DFSZ models we studied yield only fermionic ALP couplings. In this case, we found that tree-level threshold corrections can be more important and dominate the model constraints. However, the model remains weakly constrained except in regions where the UV Yukawas are large. A more dedicated study including different assumptions on the mass spectrum, additional observables or incorporating one-loop threshold corrections remains to be explored in future studies.

\section*{Acknowledgments}
We thank Mart\'in Gonz\'alez Alonso for sharing the $\chi^2$ of the global fit in \cite{Breso-Pla:2023tnz}, and the authors of \cite{Bauer:2021mvw} for providing us with a \texttt{Mathematica} version of their results. A.B.~gratefully acknowledges support from the Alexander-von-Humboldt foundation as a Feodor Lynen Fellow. J.F.M. thanks the Theoretical High Energy Physics Department at JGU Mainz for hospitality and support during his stay as a visitor. The work of J.F.M. is supported by the Spanish Ministry of Science and Innovation (MCIN) and the European Union NextGenerationEU/PRTR under grant IJC2020-043549-I, by the MCIN and State Research Agency (SRA) projects PID2019-106087GB-C22 and PID2022-139466NB-C21
(ERDF), and by the Junta de Andaluc\'ia projects P21\_00199 and FQM101. The research of A.M.G.\ and M.N.\ was supported by the Cluster of Excellence {\em Precision Physics, Fundamental Interactions and Structure of Matter\/} (PRISMA${}^+$ -- EXC~2118/1) within the German Excellence Strategy (project ID 39083149).

\appendix
\newpage
\section{Experimental inputs}
\label{app:exp_inputs}
The experimental observables included in our fit from the Higgs and top sectors are summarized in Tables~\ref{tab:obset}--\ref{tab:obset_top2}. 

\begin{table}[h]
	\centering
	\renewcommand{\arraystretch}{2.0}
    \caption{Higgs observables included in the fit.}
	\begin{adjustbox}{width=0.9\textwidth}
		\label{tab:obset}
		\begin{tabular}{|c|c|c|c|}
			\hline
			\multicolumn{2}{|c|}{Observables} & no. of measurements	 &  References \\
			\hline
			\multicolumn{2}{|c|}{\bf{Higgs Data}} & 154 & \multirow{1}{*}{}  \\ 
			\cline{1-3}
			\multirow{3}{*}{7 and 8 TeV } & ATLAS \& CMS combination  & \multirow{1}{*}{20} & \multirow{1}{*}{Table~8 of \cite{Khachatryan:2016vau}} \\
			\cline{2-4}
			\multirow{3}{*}{Run-I data }& ATLAS \& CMS combination $\mu( h \to \mu \mu)$ &  \multirow{1}{*}{1} &  \multirow{1}{*}{Table~13 of \cite{Khachatryan:2016vau}}  \\ 
            \cline{2-4}
			& ATLAS $\mu (h \to Z \gamma)$ & \multirow{1}{*}{1} & \multirow{1}{*}{ Figure~1 of \cite{Aad:2015gba}}  \\
			\hline
			\multirow{4}{*}{13 TeV ATLAS} &  $\mu ( h \to Z \gamma )$ at 139 $\ifb$ & 1 &  \cite{Aad:2020plj} \\
			& $\mu ( h \to \mu \mu)$ at 139 $\ifb$ & 1 & \cite{Aad:2020xfq} \\
			Run-II data  & $\mu(h \to \tau \tau)$ at 139 $\ifb$ & 4 & Figure~14 of \cite{ATLAS-CONF-2021-044} \\
		    & $\mu( h \to bb)$ in VBF and ${ttH}$ at 139 $\ifb$ & 1+1 & \cite{ATLAS:2020bhl,ATLAS:2020syy}  \\
		    \cline{2-4} 
		    & STXS $h \to \gamma \gamma/ZZ/b \bar{b}$ at $139\ifb$ & 42 & Figures~1 and 2 of~\cite{ATLAS:2020naq} \\
			& STXS $ h \rightarrow$ $W W$ in ggF, VBF at $139\ifb$ & 11 & Figures~12 and 14 of~\cite{ATLAS:2021upe} \\
			\hline
			 &  $\mu(h \to b \bar{b})$ in $Vh$ at $35.9/41.5\ifb$ & 2  &   entries from Table~4 of~\cite{CMS:2020gsy} \\
			 &  $\mu(h \to W W)$ in ggF at $137\ifb$ & 1  & \cite{CMS:2020dvg} \\
			 13 TeV CMS &  $\mu (h\to \mu \mu)$ at  $137\ifb$ & 4  & Figure~11 of \cite{CMS:2020xwi} \\
			Run-II data & $\mu (h \to \tau \tau/WW)$ in  $t\bar{t}h$ at $137\ifb$ & 3  & Figure~14 of~\cite{CMS:2020mpn} \\
			\cline{2-4} 
			 & STXS $h\to WW$ at $137\ifb$ in $Vh$  & 4 &  Table~9 of~\cite{CMS:2021ixs} 
			\\
			& STXS $h \to \tau \tau$ at  $137\ifb$  & 11 &  Figures~11 and 12 of~\cite{CMS:2020dvp} 
			\\
			& STXS $h \to \gamma \gamma$ at  $137\ifb$ & 27 & Table~13 and Figure~21 of~\cite{CMS:2021kom} 
			\\
			& STXS $h \to ZZ $ at  $137\ifb$ & 18 & Table~6 and Figure~15 of~\cite{CMS:2021ugl} 
			\\
			\hline
			\multicolumn{2}{|c|}{{\bf{ ATLAS $Zjj$  13 TeV $\Delta \phi_{jj}$}} at $139\ifb$}&12 & Figure~7(d) of \cite{ATLAS:2020nzk}  \\
     \hline
		\end{tabular}
	\end{adjustbox}
\end{table}

\begin{table}[ht]
	\centering
	\renewcommand{\arraystretch}{2.0}
	\caption{Top physics observables from Tevatron and LHC Run I included in the fit.}
	\begin{adjustbox}{width=0.9\textwidth}
		\label{tab:obset_top}
		\begin{tabular}{|c|c|c|c|}
			\hline
			\multicolumn{2}{|c|}{Observables} & no. of meas. &  References \\
			\hline
        \multicolumn{2}{|c|}{\bf{Top Data from Tevatron and LHC Run I}} & 82 &  \\ 
			\cline{1-3} 
 Tevatron & 
 forward-backward asymmetry $A_{FB}(m_{t\bar{t}})$ for $\mathrm{t}\overline{\mathrm{t}}$ production  &
$4$ &
 ~\cite{CDF:2017cvy} \\ \hline
\multirow{2}{*}{ATLAS \& CMS}
    & 
    charge asymmetry $A_C(m_{t\bar{t}})$ for $\mathrm{t}\overline{\mathrm{t}}$ production in the $\ell$+jets channel.  &
 $6$ &
 ~\cite{ATLAS:2017gkv} \\ \cline{2-4} 
 & 
   $W$-boson helicity fractions in top decay 
 & $3$ &  ~\cite{CMS:2020ezf} \\ 
  \hline
 \multirow{8}{*}{ATLAS} &  charge asymmetry $A_C(m_{t\bar{t}})$ for $\mathrm{t}\overline{\mathrm{t}}$ production in the dilepton channel  &
 $1$ &
 ~\cite{ATLAS:2016ykb} \\  \cline{2-4} 
     & $\sigma_{t\bar{t}W}, \, \sigma_{t\bar{t}Z}$ &
 $2$ &
 ~\cite{ATLAS:2015qtq} \\  \cline{2-4} 
   & $\tfrac{d\sigma}{dp^T_{t}} ,\quad \tfrac{d\sigma}{d|y_{\bar{t}}}$  for $t$-channel single-top production
 &
 $4+5$ &
 ~\cite{ATLAS:2017rso} \\ \cline{2-4} 
  &  $\sigma_{tW}$ in the single lepton channel &
 $1$ & ~\cite{ATLAS:2020cwj} \\ \cline{2-4} 
 & $\sigma_{tW}$ in the dilepton channel &
 $1$ & ~\cite{ATLAS:2015igu} \\ \cline{2-4} 
& $s$-channel single-top cross section &
 $1$ & ~\cite{ATLAS:2015jmq} \\ \cline{2-4} 
& $\tfrac{d\sigma}{dm_{t\bar{t}}}$ for $t\bar{t}$ production in the dilepton channel
  &
 $6$ &
 ~\cite{ATLAS:2016pal} \\ \cline{2-4} 
 & $\tfrac{d\sigma}{dp^T_{t}}$  for $t\bar{t}$ production in the $\ell$+jets channel
 & $8$ &
 ~\cite{ATLAS:2015lsn} \\ \hline
  \multirow{10}{*}{CMS}    & $\sigma_{t\bar{t}\gamma}$ in the $\ell+$ jets channel.&
 $1$ &
 ~\cite{CMS:2015uvn} \\ \cline{2-4} 
  & charge asymmetry $A_C(m_{t\bar{t}})$ for $\mathrm{t}\overline{\mathrm{t}}$ production in the dilepton channel.  &
 $3$ &
 ~\cite{CMS:2016ypc} \\ \cline{2-4} 
  & $\sigma_{t\bar{t}W}, \, \sigma_{t\bar{t}Z}$ &
 $2$ &
  ~\cite{CMS:2015uvn} \\ \cline{2-4} 
   & $\sigma_{t\bar{t}\gamma}$ in the $\ell+$ jets channel.&
 $1$ &
  ~\cite{CMS:2017tzb} \\ \cline{2-4} 
   & $s$-channel single-top cross section &
 $1$ &
 ~\cite{CMS:2016xoq} \\ \cline{2-4} 
  &   $\tfrac{d\sigma}{dp^T_{t+\bar{t}}}$ of $t$-channel single-top production
 &
 $6$ & ~\cite{CMS:2014ika} \\ \cline{2-4} 
  & $t$-channel single-top and anti-top cross sections $R_t$. &
 $1$ & ~\cite{CMS:2014mgj} \\ \cline{2-4} 
  & $\sigma_{tW}$ &
 $1$ & ~\cite{CMS:2014fut} \\ \cline{2-4} 
   & $\tfrac{d\sigma}{dm_{t\bar{t}}dy_{t\bar{t}}}$ for $t\bar{t}$ production in the dilepton channel
  & $16$ &
 ~\cite{CMS:2017iqf,CMS:2013hon} \\ \cline{2-4} 
  &  $\tfrac{d\sigma}{dp^T_{t}}$ for $t\bar{t}$ production in the $\ell$+jets channel
  & $8$ &
 ~\cite{CMS:2015rld,CMS:2016csa} \\ \hline
		\end{tabular}
	\end{adjustbox}
\end{table}
 
\begin{table}[ht]
	\centering
	\renewcommand{\arraystretch}{2.0}
	\caption{Top physics observables from LHC Run~II included in the fit.}
	\begin{adjustbox}{width=0.9\textwidth}
		\label{tab:obset_top2}
		\begin{tabular}{|c|c|c|c|}
			\hline
			\multicolumn{2}{|c|}{Observables} & no. of meas. &  References \\
			\hline
        \multicolumn{2}{|c|}{\bf{Top Data from LHC Run II}} & 55 &  \\ 
			\cline{1-3} 
  \multirow{6}{*}{ATLAS} & $\sigma_{tW}$  &
 $1$ &
 ~\cite{ATLAS:2016ofl} \\ \cline{2-4} 
   & $\sigma_{tZ}$ &
 $1$ &
 ~\cite{ATLAS:2017dsm} \\ \cline{2-4} 
  & $\sigma_{t+\bar{t}}, \, R_t$ for $t$-channel single-top and anti-top cross sections&
 1+1 &~\cite{ATLAS:2016qhd} \\ \cline{2-4} 
   &   charge asymmetry $A_C(m_{t\bar{t}})$ for $\mathrm{t}\overline{\mathrm{t}}$ production &
 $5$ &
 ~\cite{ATLAS:2019czt} \\ \cline{2-4} 
  &  $\sigma_{t\bar{t}W}, \, \sigma_{t\bar{t}Z}$  &
 $2$ &
 ~\cite{ATLAS:2019fwo} \\ \cline{2-4} 
   &  $\tfrac{d\sigma}{dp^T_{\gamma}}$ for  $t\bar{t}\gamma$ production
 &
 $11$ &
 ~\cite{ATLAS:2020yrp} \\ \hline
  \multirow{7}{*}{CMS} & $\sigma_{tW}$ &
 $1$ &
 ~\cite{CMS:2018amb} \\ \cline{2-4} 
   & $\sigma_{tZ}$ in the $Z\to\ell^+\ell^-$ channel  &
 $1$ &
 ~\cite{CMS:2018sgc} \\ \cline{2-4} 
 & 
 $\tfrac{d\sigma}{dp^T_{t+\bar{t}}}$ and $R_t\left(p^T_{t+\bar{t}}\right)$ for $t$-channel single-top quark production &
 $5+5$ &
 ~\cite{CMS:2019jjp} \\ \cline{2-4} 
   & $\tfrac{d\sigma}{dm_{t\bar{t}}}$  for $t\bar{t}$ production in the dilepton channel
 &
 $6$ &
 ~\cite{CMS:2018fks}  \\ \cline{2-4} 
 & $\tfrac{d\sigma}{dm_{t\bar{t}}}$ for $t\bar{t}$ production in the $\ell+$jets channel
&
 $15$ &
 \cite{CMS:2021fhl}  \\ \cline{2-4} 
 & $\sigma_{t\bar{t}W}$ &
 $1$ &
 ~\cite{CMS:2017ugv} \\ \cline{2-4} 
 & $\tfrac{d\sigma}{dp^T_{Z}}$ for $t\bar{t}Z$ production  &
 $4$ &
 ~\cite{CMS:2019too} \\ \hline
		\end{tabular}
	\end{adjustbox}
\end{table}

\FloatBarrier
\section{Contributions to \texorpdfstring{$\bm{Z}$}{Z}-pole observables in the LL approximation}
\label{sec:LE_Expressions}
In our global analysis, we assume flavor-universal ALP couplings. Here we provide the expressions for the SMEFT Wilson coefficients in the Warsaw basis~\cite{Grzadkowski:2010es} which enter $Z$-pole observables in the LL approximation in terms of the ALP coefficients in \eqref{eq:lag2}. Keeping only the entries $[\bm{Y}_u]_{33}\equiv y_t$, $[\bm{Y}_d]_{33}\equiv y_b$ and $[\bm{Y}_e]_{33}\equiv y_\tau$, the SMEFT Wilson coefficients at $\mu=m_Z$ read
\begin{align} 
C_{HWB}&=4\,g_L\,g_Y\, C_{BB}\, C_{WW}\, \ln\frac{\Lambda}{m_Z}\,,\notag\\
C_{HD}&=-\frac{8}{3}\,g_Y^2\, C_{BB}^2\, \ln\frac{\Lambda}{m_Z}\,,\notag\\
\big[C_{Hq}^{(1)}\big]_{ij}&=\left[-\frac{4}{9}\,g_Y^2\,C_{BB}^2\,\delta_{ij}+\frac{1}{4}\,\left(y_t^2\,C_u^2 - y_b^2\,C_d^2\right) \delta_{i3}\delta_{j3}\right]\ln\frac{\Lambda}{m_Z}\,,\notag\\
\big[C_{Hq}^{(3)}\big]_{ij}&=\left[-\frac{4}{3}\,g_L^2\,C_{WW}^2\,\delta_{ij}-\frac{1}{4}\,\left(y_t^2\,C_u^2 + y_b^2\,C_d^2 \right) \delta_{i3}\delta_{j3}\right] \ln\frac{\Lambda}{m_Z}\,,\notag\\
\big[C_{Hu}\big]_{ij}&=\left(-\frac{16}{9}\,g_Y^2\,C_{BB}^2\,\delta_{ij}-\frac{1}{2}\,y_t^2\,C_u^2\,\delta_{i3}\delta_{j3}\right)\ln\frac{\Lambda}{m_Z}\,,\notag\\
\big[C_{Hd}\big]_{ij}&= \left(\frac{8}{9}\,g_Y^2\,C_{BB}^2\,\delta_{ij}+\frac{1}{2}\,y_b^2\,C_d^2\,\delta_{i3}\delta_{j3} \right)\ln\frac{\Lambda}{m_Z}\,,\notag\\
\big[C_{Hud}\big]_{ij}&=y_b\,y_t\,C_d\,C_u\,\delta_{i3}\delta_{j3}\, \ln\frac{\Lambda}{m_Z}\,,\notag\\
\big[C_{Hl}^{(1)}\big]_{ij}&=\left(\frac{4}{3}\,g_Y^2 \,C_{BB}^2\,\delta_{ij} - \frac{1}{4}\,y_\tau^2\,C_e^2\,\delta_{i3}\delta_{j3} \right)\ln\frac{\Lambda}{m_Z}\,,\notag\\
\big[C_{Hl}^{(3)}\big]_{ij}&=\left(-\frac{4}{3}\,g_L^2\,C_{WW}^2\,\delta_{ij} - \frac{1}{4}\,y_\tau^2\,C_e^2\,\delta_{i3}\delta_{j3}\right)\ln\frac{\Lambda}{m_Z}\,,\notag\\
\big[C_{He}\big]_{ij}&=\left(\frac{8}{3}\,g_Y^2\,C_{BB}^2\,\delta_{ij} + \frac{1}{2}\,y_\tau^2\,C_e^2 \,\delta_{i3}\delta_{j3}\right)\ln\frac{\Lambda}{m_Z}\,,\notag\\
\big[C_{ll}\big]_{1221}&=-\frac{4}{3}\,g_L^2\,C_{WW}^2\,\ln\frac{\Lambda}{m_Z}\,,
\end{align}
where again $\Lambda=4\pi f$, and the SMEFT Wilson coefficients normalized as $C_i/\Lambda^2$. Parametrizing the $Z$ and $W$ coupling modifications as in~\cite{Falkowski:2017pss}, we obtain the following relations:
\begin{align}
\delta m_W^2&=\frac{v^4}{\Lambda^2}\,\frac{g_Y^2\,g_L^2}{g_L^2-g_Y^2}\left( \frac{2}{3}\,C_{BB}^2 - 4\, C_{BB}\,C_{WW} + C_{WW}^2 \right)\ln\frac{\Lambda}{m_Z}\,,\notag\\
[\delta g_R^{Wq} ]_{ij} &= -\frac{v^2}{\Lambda^2}\,\frac{y_b\,y_t}{2}\,C_d\,C_u\,\delta_{i3}\delta_{j3}\,\ln\frac{\Lambda}{m_Z}\,,\notag\\
[\delta g_L^{Zu} ]_{ij} &= \frac{v^2}{\Lambda^2}\,\bigg[\frac{\delta_{ij}}{g_L^2 - g_Y^2}\,\bigg( \frac{1}{9} \,g_Y^2\,(5g_L^2-g_Y^2)\,C_{BB}^2 - \frac{8}{3}\,g_L^2\,g_Y^2\,C_{BB}\,C_{WW} - \frac{1}{6}\,g_L^2\,(g_L^2 - 5g_Y^2)\,C_{WW}^2 \bigg)\notag\\
&\hspace{1.5cm}  - \frac{\delta_{i3}\delta_{j3}}{4}\,y_t^2\,C_u^2\bigg]\ln\frac{\Lambda}{m_Z}\,,\notag\\
[\delta g_L^{Zd} ]_{ij} &= \frac{v^2}{\Lambda^2}\bigg[ \frac{\delta_{ij}}{g_L^2 - g_Y^2}\,\left(-\frac{1}{9}\,g_Y^2 (g_L^2 + g_Y^2) C_{BB}^2 + \frac{4}{3}\,g_L^2\,g_Y^2 \,C_{BB}\,C_{WW} + \frac{1}{6}\,g_L^2(g_L^2 - 3g_Y^2)C_{WW}^2\right)\notag\\
&\hspace{1.5cm}  + \frac{\delta_{i3}\delta_{j3}}{4}\,y_b^2\,C_d^2\bigg]\ln\frac{\Lambda}{m_Z}\,,\notag\\
[\delta g_R^{Zu} ]_{ij} &= \frac{v^2}{\Lambda^2}\,\bigg[\frac{g_Y^2}{g_L^2 - g_Y^2}\,\delta_{ij}\,\bigg( \frac{4}{9}\,(2g_L^2-g_Y^2)\,C_{BB}^2 - \frac{8}{3}\,g_L^2\,C_{BB}\,C_{WW} + \frac{2}{3}\,g_L^2\,C_{WW}^2\bigg)\notag\\
&\hspace{1.5cm} + \frac{\delta_{i3}\delta_{j3}}{4}\,y_t^2\,C_u^2\bigg]\ln\frac{\Lambda}{m_Z}\,,\notag\\
[\delta g_R^{Zd} ]_{ij} &= \frac{v^2}{\Lambda^2}\bigg[\frac{g_Y^2 }{g_L^2 - g_Y^2}\,\delta_{ij}\,\bigg(\frac{2}{9}\,(g_Y^2-2g_L^2)\,C_{BB}^2+\frac{4}{3}\, g_L^2 \,C_{BB}\,C_{WW} - \frac{1}{3}\,g_L^2\,C_{WW}^2\bigg)\notag\\
&\hspace{1.5cm} - \frac{\delta_{i3}\delta_{j3}}{4}\,y_b^2\,C_d^2\bigg]\ln\frac{\Lambda}{m_Z}\,,\notag\\
[\delta g_L^{Wl} ]_{ij} &= \frac{v^2}{\Lambda^2}\bigg[\frac{g_L^2 }{g_L^2 - g_Y^2}\,\delta_{ij}\left(\frac{2}{3}\,g_Y^2\,C_{BB}^2 - 4\,g_Y^2\,C_{BB}\,C_{WW} - \frac{1}{3}\,(g_L^2 - 4\,g_Y^2)\,C_{WW}^2\right)\notag\\
&\hspace{1.5cm} - \frac{\delta_{i3}\delta_{j3}}{4}\,y_\tau^2\,C_e^2\bigg]\ln\frac{\Lambda}{m_Z}\,,\notag\\
[\delta g_L^{Ze} ]_{ij} &= \frac{v^2}{\Lambda^2}\bigg[\frac{\delta_{ij} }{g_L^2 - g_Y^2}\bigg( \frac{1}{3}\,g_Y^2(g_Y^2 - 3 g_L^2)\,C_{BB}^2 + 4\,g_L^2\,g_Y^2\,C_{BB}\,C_{WW} + \frac{1}{6}g_L^2(g_L^2 - 7 g_Y^2 )\,C_{WW}^2\bigg)\notag\\
&\hspace{1.5cm} + \frac{\delta_{i3}\delta_{j3}}{4}\,y_\tau^2\,C_e^2\bigg]\ln\frac{\Lambda}{m_Z}\,,\notag\\
[\delta g_R^{Ze} ]_{ij} &= \frac{v^2}{\Lambda^2}\bigg[\frac{g_Y^2}{g_L^2 - g_Y^2}\,\delta_{ij} \,\left(\frac{2}{3}(g_Y^2-2g_L^2)\,C_{BB}^2 + 4\,g_L^2\,C_{BB}\,C_{WW} - g_L^2\,C_{WW}^2\right)\notag\\
&\hspace{1.5cm} - \frac{\delta_{i3}\delta_{j3}}{4}\,y_\tau^2\,C_e^2\bigg]\ln\frac{\Lambda}{m_Z}\,,
\end{align}
with the additional relations $\delta g_L^{Z\nu}=\delta g_L^{Ze}+\delta g_L^{Wl}$ and $\delta g_L^{Wq}\approx\delta g_L^{Zu}-\delta g_L^{Zd}$ (for $V_{\rm CKM}\approx\mathbb{1}$). From these expressions it becomes clear why $C_u$ remains unconstrained by $Z$-pole observables when working at LL accuracy, as this parameter only enters in $Zt\bar t$ coupling modifications.

\newpage
\bibliographystyle{JHEP}
\bibliography{references}

\end{document}